\documentclass[twocolumn,showpacs,a4,prl,superscriptaddress]{revtex4-2} 
\usepackage{graphicx}%
\usepackage{color}
\usepackage{dcolumn} 
\usepackage{amsmath} 
\usepackage[utf8x]{inputenc} 
\usepackage{amssymb} 
\usepackage{bm} 
\usepackage{hyperref}
\usepackage{gensymb}
\usepackage{mathtools}
\usepackage{amsmath}
\usepackage{amssymb}
\usepackage{amsthm}
\usepackage{amsfonts}
\usepackage{braket}
\usepackage[outdir=./]{epstopdf}
\usepackage{tabularx}
\hypersetup{colorlinks,%
						citecolor=blue,%
						filecolor=blue,%
						linkcolor=blue,%
						urlcolor=blue,%
						pdftex}
\fontfamily{ptm}
\begin{document}
\preprint{WTe2 NMR.TEX}  
\title{Detection of Weyl Fermions and the Metal to Weyl-Semimetal phase transition in WTe$_2$ via broadband High Resolution NMR}

\author{W. Papawassiliou}
\affiliation{Department of Materials and Environmental Chemistry, Arrhenius Laboratory, Stockholm University, Svante Arrhenius väg 16 C, SE-106 91 Stockholm, Sweden}
\author{J.P. Carvalho}
\affiliation{Department of Materials and Environmental Chemistry, Arrhenius Laboratory, Stockholm University, Svante Arrhenius väg 16 C, SE-106 91 Stockholm, Sweden}
\author{H.J. Kim}
\email[Corresponding author: ]{hansol@re.kbsi.kr}
\affiliation{Research Center for Materials Analysis, Korea Basic Science Institute, 169-148 Gwahak-ro, Yuseong-gu, Daejeon 34133, Republic of Korea}
\author{C.Y. Kim}
\affiliation{Research Center for Materials Analysis, Korea Basic Science Institute, 169-148 Gwahak-ro, Yuseong-gu, Daejeon 34133, Republic of Korea}
\author{S.J. Yoo}
\affiliation{Research Center for Materials Analysis, Korea Basic Science Institute, 169-148 Gwahak-ro, Yuseong-gu, Daejeon 34133, Republic of Korea}
\author{J.B. Lee}
\affiliation{Research Center for Materials Analysis, Korea Basic Science Institute, 169-148 Gwahak-ro, Yuseong-gu, Daejeon 34133, Republic of Korea}
\author{S. Alhassan}
\affiliation{Department of Chemical Engineering, Khalifa University, PO Box 2533, Abu Dhabi, United Arab Emirates}
\author{S. Orfanidis}
\affiliation{Institute of Nanoscience and Nanotechnology, NCSR Demokritos, 15310 Aghia Paraskevi, Attiki, Greece}
\author{V.Psycharis}
\affiliation{Institute of Nanoscience and Nanotechnology, NCSR Demokritos, 15310 Aghia Paraskevi, Attiki, Greece}
\author{M. Karagianni}
\affiliation{Institute of Nanoscience and Nanotechnology, NCSR Demokritos, 15310 Aghia Paraskevi, Attiki, Greece}
\author{M. Fardis}
\affiliation{Institute of Nanoscience and Nanotechnology, NCSR Demokritos, 15310 Aghia Paraskevi, Attiki, Greece}
\author{N. Panopoulos}
\affiliation{Institute of Nanoscience and Nanotechnology, NCSR Demokritos, 15310 Aghia Paraskevi, Attiki, Greece}
\author{G. Papavassiliou}
\email[Corresponding author: ]{g.papavassiliou@inn.demokritos.gr}
\affiliation{Institute of Nanoscience and Nanotechnology, NCSR Demokritos, 15310 Aghia Paraskevi, Attiki, Greece}
\author{A.J. Pell}
\email[Corresponding author: ]{andrew.pell@ens-lyon.fr}
\affiliation{Centre de RMN à Très Hauts Champs de Lyon (UMR 5280 CNRS/ENS Lyon/Université Claude Bernard Lyon 1), Université de Lyon, 5 rue de la Doua, 69100 Villeurbanne, France}

\date{\today}

\begin{abstract}
 Weyl Fermions (WFs)  in the type-II Weyl Semimetal  (WSM) WTe$_2$ are difficult to resolve experimentally because the Weyl bands disperse in an extremely narrow region of the (E-k) space. Here, by using DFT-assisted high-resolution $^{125}$Te solid-state NMR (ssNMR) in the temperature range $50$K - $700$K,  we succeeded in detecting low energy WF excitations and monitor their evolution with temperature. Remarkably, WFs appear to emerge at T$\sim 120$K; at lower temperatures WTe$_2$ behaves as a metal. This intriguing metal-to-WSM phase transition is shown to be induced by the rapid raise of the Fermi level with temperature, crossing solely the electron and hole pockets in the low-T metallic phase, while crossing the Weyl bands near the nodal points - a prerequisite for the emergence of WFs - only for T$>120$K. 
\end{abstract}
\maketitle

In 1929, Hermann Weyl predicted a new type of fermion with zero mass and explicit chirality that became known as the “Weyl Fermion” (WF)\cite{Weyl1929}. For many years neutrinos were considered a unique paradigm of WFs, until it was discovered that they have mass. Since then, none of the elementary particles in high-energy physics has been identified a WF. However, recently a condensed matter analogue of this elusive particle was predicted\cite{Xu2015,Lv2015,Weng2015} in a new class of topological materials, the type-I WSMs. In these materials, space inversion symmetry (SIS) or time reversal symmetry (TRS) is broken, which leads to energy bands described by the Weyl equation, crossing linearly in pairs of nodal points with opposite chirality. Quasiparticles in the vicinity of these nodes  are emergent WFs. 

Shortly thereafter, a second class of WSMs was predicted, violating in addition to SIS and TRS the Lorentz symmetry (LS)\cite{Soluyanov2015}. In the standard model of high-energy physics, WFs are considered to obey strictly the LS. However, in theories incorporating gravity, LS is violated and the light cones become strongly tilted\cite{Volovik2017}. An analogous violation of the LS is observed in specific WSMs, revealing a novel type-II class of WSMs, with tilted Weyl cones, and a Fermi Surface (FS) comprising electron and hole pockets touching at the nodal points \cite{Soluyanov2015,Wang2016,Singh2020}.

 Until now, WFs have been experimentally observed in type-I WSMs\cite{Xu2015,Lv2015,Weng2015}, whereas in type-II WSMs they have been indirectly claimed through observation of the surface FS arcs\cite{Bruno2016,Wu2016}. The primary difficulty lies in the fact that in most type-II WSMs the Weyl nodes are located above the Fermi level E$_F$, while crossing Weyl bands disperse in an extremely narrow area in the (E-k) space\cite{Soluyanov2015,Wang2016,Singh2020}, and are thus impossible to resolve with Angle Resolved Photoemission Spectroscopy (ARPES)\cite{Hasan2021}. 

An alternative method is to use solid state NMR (ssNMR) in order to observe WFs as elementary excitations. When E$_F$ crosses the linear Weyl bands,  emergent WFs force the NMR spin-lattice relaxation time $T_1$ to vary with temperature as $\frac{1}{{T{_1}T}}$ $\sim T^{2}$$ln(\frac{k_BT}{\hbar \omega_L})$, ($\omega$${_L}$ is the Larmor frequency)\cite{Okvatovity2016,Okvatovity2019}. This behaviour was verified experimentaly in the TaP and TaAs type-I WSMs\cite{Yasuoka2017,Wang2020}. However, in type-II WSMs, where the tilted Weyl cones create a complex FS changing dramatically with temperature\cite{Wu2015,He2019,Wang2017,Pletikosic2014}, and the energy bands and the Fermi velocity $\upsilon{_F}$ might be renormalized\cite{Hirata2017,Hirata2016}, the strong coupling between the spin and orbital degrees of freedom obscures the experimental verification of WFs.  It is thus necessary to find ways to resolve FS electron-spin excitations from WF orbital excitations. Bearing these things in mind, the type-II WSM WTe$_2$ was studied by combining advanced broadband $^{125}$Te ssNMR techniques with density functional theory (DFT) calculations of both the electron energy bands and the NMR Knight shift $K$ parameters to precisely tie individual NMR resonances with the Weyl electron states via the electron-nuclear hyperfine coupling. 

\begin{figure}[h]
	\includegraphics[width=0.5\textwidth]{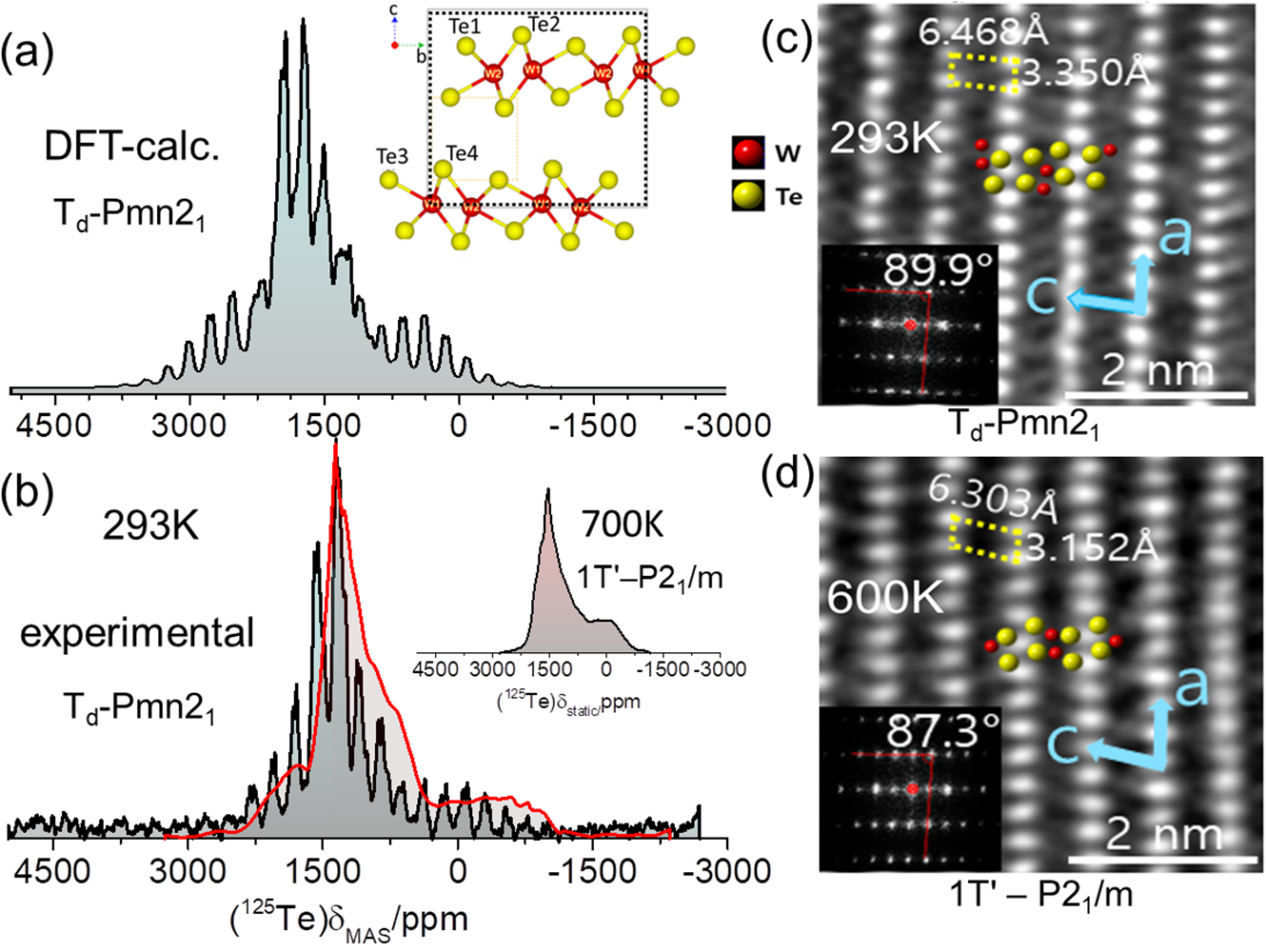}
\caption{\label{Fig1} (a,b) experimental and DFT calculated 1-D $^{125}$Te MAS NMR spectra (black lines) of WTe$_2$ in the T$_d$ phase at 293K in external magnetic field $9.4$T. The red line is the relevant static NMR spectrum. The inset shows the  $^{125}$Te static NMR spectrum in the 1T$^\prime$ phase at 700K. (c,d)  HRTEM images vertical to the $<010>$ zonal axis at temperatures 293K and 600K. The insets display the relevant EDPs. A change of the (002)/(202)/(200) intersection angle from 90$^{\circ}$ to 87.3$^{\circ}$ marks the T$_d$ $\rightarrow$ 1T$^\prime$ phase transition. }
	\end{figure}

WTe$_2$ crystallizes in two different crystal phases; at low temperatures it crystallizes in the T$_d$ phase\cite{Dawson1987}  belonging to the non-centrosymmetric orthorhombic space group Pmn2$_{1}$, while upon heating to $\sim$ 550K, it undergoes a structural phase transition to the 1T$^\prime$ phase\cite{Tao2020}  with the centrosymmetric monoclinic space group P2$_1$/m. The primary difference between the two phases is that in the 1T$^\prime$ phase the c-axis is inclined by an angle of 2.7$^{\circ}$ from the vertical position, as shown by the high-resolution transmission microscope (HRTEM) images in Figures~\ref{Fig1}c,d, and Supplementary Figures 2-5.  The lack of inversion symmetry in the Pmn2$_{1}$ space group is the key parameter responsible for the topologically non-trivial character of the electron energy bands of the T$_d$ phase. 
\indent
\indent

Figures ~\ref{Fig1}a,b show the experimental and DFT calculated 1-D $^{125}$Te MAS (black lines) and static (red line) ssNMR spectra of WTe$_2$ in the T$_d$ phase at $293$K. Experimental spectra span over the frequency range of 2500 ppm to -1000 ppm, exhibiting a broad asymmetric peak with a maximum at $\sim$1350 ppm; it is however, impossible to deconvolute them into signal components assigned to the four inequivalent Te sites. Notably, a nice agreement is observed between the experimental and DFT calculated spectra, which in principle allows the 1-D NMR spectrum deconvolution, however without any direct experimental confirmation.Details on the DFT calculations are presented below and in the Suplementary Information. Upon heating up to 700K, a significant decrease in the anisotropy of the spectrum is observed by a narrowing of the resonance and a shift of the main peak to 1650 ppm, as shown in the inset of Figure ~\ref{Fig1}b.     

\begin{figure*}[t]
	\includegraphics[width=0.8\textwidth]{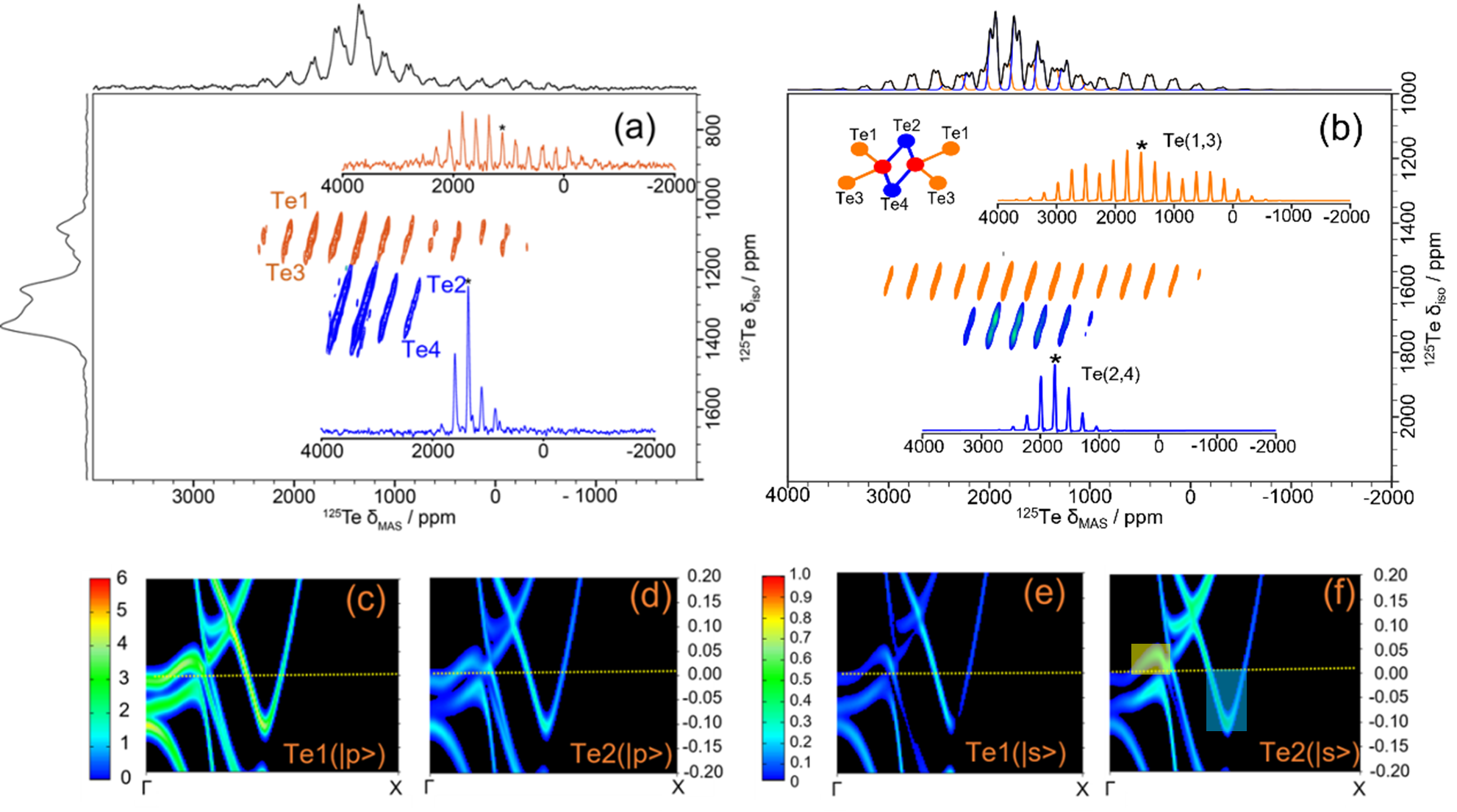}
	\caption{\label{Fig2} (a) Experimental 2-D $^{125}$Te aMAT NMR spectrum at 30 kHz MAS in magnetic field of 9.4T. An equivalent spectrum at 14.1T is presented in Supplementary Figure 9. (b) DFT calculated 2-D $^{125}$Te aMAT NMR spectrum in the presence of SOC. (c-f) The relevant Te(1) and Te(2) $\ket{s}$ and $\ket{p}$ orbitals contribution to the k-resolved pDOS. Hole and electron pockets are highlighted in yellow, respectively blue color, in panel (f).}
	\end{figure*}
\indent
To break down the 1-D NMR lineshape into the multiple signals representing the various Te environments we performed a 2-D adiabatic magic angle turning (aMAT)\cite{Papawassiliou2020,Clement2012} experiment (Figure~\ref{Fig2}a), which allows to separate the isotropic NMR Knight shift $K_{iso}$ from the Knight shift anisotropy $\Delta$. In the resulting spectrum, two distinct groups of shifts are observed, one group having isotropic Knight shifts $K_{iso}$ = 1080/1128  ppm and Knight shift anisotropies $\Delta$ = -2305.8/-2861.5 ppm assigned to Te(1)/Te(3) respectively, and the second with $K_{iso}$  = 1269/1351 ppm and $\Delta$ = -855.51/-786.30  ppm for Te(2) and Te(4). The assignment of the shifts to the distinct Te sites was based on the DFT calculations of the Knight shifts, performed within the formalism of the WIEN2k package. \indent

In general, $K$ can be broken down into three major components, according to $K$=$K_{orb}$+$K_{FC}$+$K_{dip}$, where $K_{orb}$ is the orbital term, $K_{FC}$ the Fermi contact term, and K$_{dip}$ the dipolar term\cite{Papawassiliou2020,Laskowski2015}. In the case of metallic systems, to a first non-interacting electrons approximation, $K_{FC}$ is dominant and proportional to the s-electrons projected density of states (pDOS) at E$_F$\cite{Papawassiliou2021,Korringa1950,Knight1949}. However, in topologically non-trivial systems, in the presence of strong SOC, the main component of the Knight shift is $K_{orb}$ \cite{Papawassiliou2020,Okvatovity2016,Okvatovity2019,Maebashi2019}, induced mainly by the interaction of the p-electron orbital currents with the nuclear spins. The DFT calculated aMAT spectrum is presented in Figure~\ref{Fig2}b, exhibiting remarkable consistency with the experimental one; however, an overall frequency shift is observed, due to the sensitivity of the NMR calculations to the final atomic positions in the DFT-relaxed structure. All information regarding the experimentally derived and calculated NMR parameters is presented in Supplementary Table 1. The key outcome is that in the case of Te(1) and Te(3), $K_{orb}$ is more than an order of magnitude larger than $K_{FC}$, whilst for Te(2), Te(4) the $K_{FC}$ has a considerable contribution, being smaller than $K_{orb}$ only by a factor of two. For all Te sites, $K_{dip}$ is found to be negligibly small. On the basis of this analysis, further DFT simulations of 1D $^{125}$Te MAS NMR spectra at spinning rate 30 kHz and magnetic fields of 9.4, 14.1, and 23.5T, were carried out and compared with experimental $^{125}$Te MAS NMR spectra, acquired by a rotor-synchronized double adiabatic echo pulse sequence (Supplementary Figures 7 and 8).  The agreement between the theoretical and experimental spectra validates the correctness of our approach.
\indent 

Figures~\ref{Fig2}c-f show the DFT calculated s- and p-orbitals k-resolved pDOS of the Te(1) and Te(2) sites, obtained in the same set of calculations as the theoretical NMR spectra. The well-known WTe$_2$ band structure is observed, with  hole and electron pockets, highlighted in yellow, respectively blue color. The k-pDOS of Te(1) acquires a strong p-orbital character while Te(2) exhibits a mixed s- and p-orbitals character in line with the NMR DFT analysis of Supplementary Table 1 and the pDOS values presented in Supplementary Figure 10. Evidently, the strong orbital anisotropy of the Te(1) NMR signal component is driven by the prevalent p-orbital character of the relevant pDOS and Te(1)/Te(3) NMR stands out as an excellent probe of low energy WF excitations. At the same time Te(2)/Te(4) NMR is susceptible to both s-electron spin and p-electron orbital fluctuations.

\begin{figure*}[t]
	\includegraphics[width=0.95\textwidth]{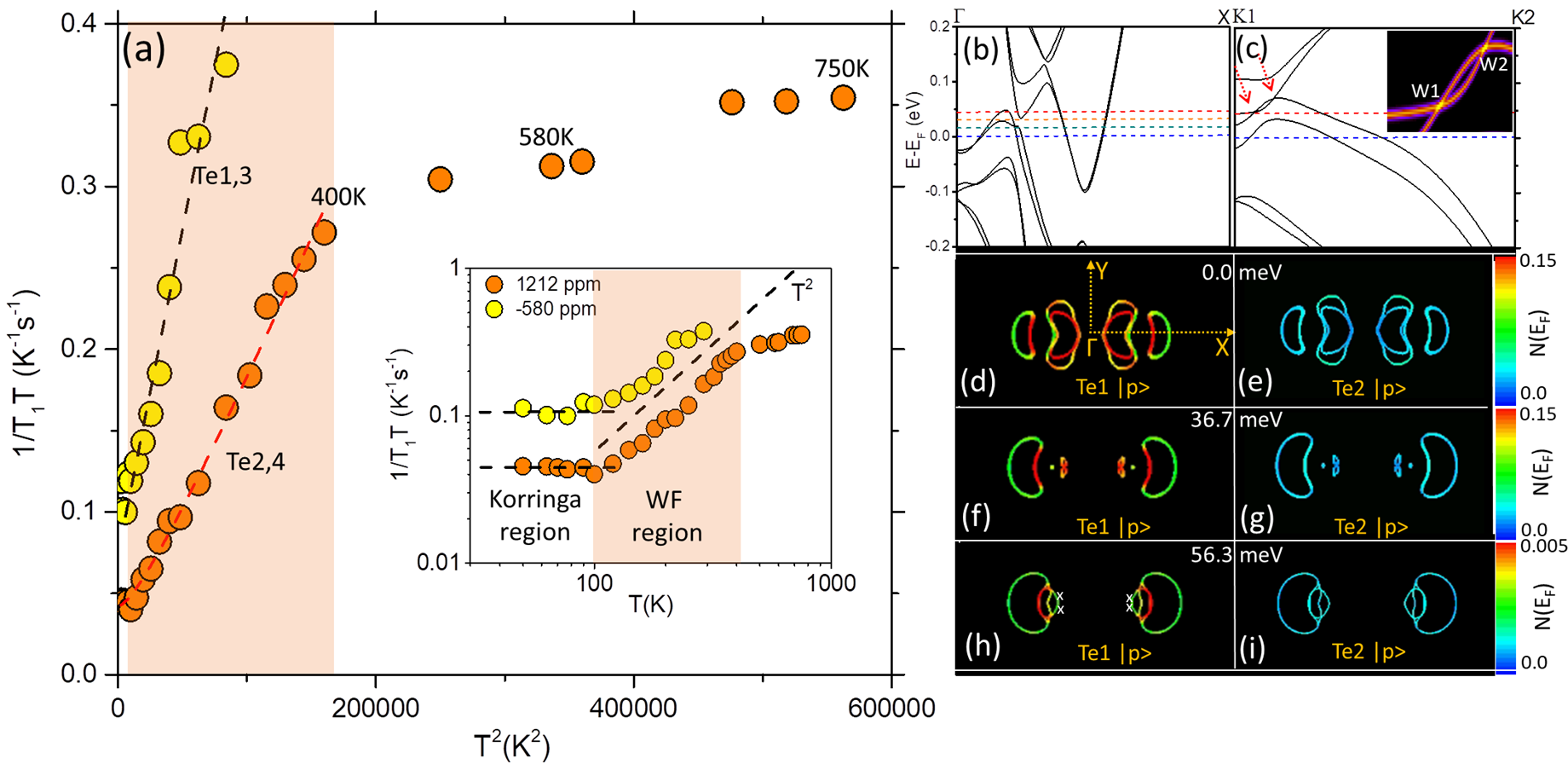}
	\caption{\label{Fig3} (a) $^{125}$Te $\frac{1}{{T{_1}T}}$ vs. $T^{2}$ plot, acquired at -580 ppm (yellow circles) and 1212 ppm (orange circles). (b-c) The band structure of the T$_d$ phase along the lines connecting the $\Gamma$-X high symmetry points and the K1=(0.1212, 0.0000, 0.0000) --  K2=(0.0929, 0.2000, 0.0000) points. A pair of conjugate Weyl points (W1=0.119,0.020,0.000) and (W2= 0.117,0.034,0.000) is observed along the K1-K2 direction. The color dashed lines indicate the raise of E$_F$ with temperature\cite{Wu2015,He2019,Wang2017,Pletikosic2014}. (d-i) FS cross-sections of Te(1)/Te(2)  p-electron bands, at k$_{z}$=0, and E$_F$= 0.0, 36.7 meV, and 56.3 meV. White crosses “x” in panel (h) indicate the positions of pairs of Weyl points.}
	\end{figure*}
Taking into account the previous results, the spin-lattice relaxation time $T_{1}$ was measured over the temperature range of 50K – 750K at two distinct frequency shifts on the inhomogeneous static NMR lineshape (red line in Figure~\ref{Fig1}b), i.e. at 1212 ppm probing mainly Te(2), Te(4) nuclear sites, and -580 ppm probing Te(1) and Te(3). Figure~\ref{Fig3}a presents the $\frac{1}{{T{_1}T}}$ vs. T$^{2}$  variation at both shifts. At temperatures below 120K the plot sampled at 1212 ppm is flat, obeying a Korringa-like ($\frac{1}{{T{_1}T}}$ = ct) relation, which is characteristic of metals\cite{Korringa1950}; then it experiences an uptick at $\sim$120 $K$, followed by a $T^{2}$ temperature dependence, which is literally assigned to the formation of WFs \cite{Okvatovity2016,Okvatovity2019,Yasuoka2017,Wang2020,Maebashi2019}. A similar behavior is observed in the $\frac{1}{{T{_1}T}}$ vs.  T$^{2}$ plot sampled at -580 ppm; however, the slope in the WF region is sufficiently steeper than the one sampled at 1212 ppm.  The slope difference might be explained by the fact that at -580 ppm, NMR effectively probes almost purely orbital excitations, while at 1212 ppm orbital and FS electron spin excitations are intermixed.  
The detected transition from the low-T metallic regime to the WF phase is of primary importance for understanding the Weyl Physics of WTe$_2$. In conventional metals, the Fermi level does not change appreciably for k$_{B}$T$\ll$E$_{F}$ and $\frac{1}{{T{_1}T}}=$ct\cite{Papawassiliou2020}. However, in WTe$_{2}$, where the hole pocket maximum and the electron pocket minimum are in proximity to E$_{F}$, a significant raise of E$_{F}$ with temperature has been reported\cite{Wu2015,He2019,Wang2017,Pletikosic2014}, inducing shrinkage of the hole pockets and a topological Lifshitz transition (LT)\cite{Lifshitz1960} taking place at $\sim$160K\cite{Wu2015}, which is associated with the disappearance of the hole pockets\cite{Wu2015,He2019}. Specifically, ARPES experiments show an E$_{F}$ rise of 30 meV in the temperature range 20K - 100K\cite{Pletikosic2014}, and 25 meV between 100K - 280K\cite{Wu2015}.   

The transition from the low-T metallic to the high-T WF phase is highlighted in the DFT band structure calculations in Figures~\ref{Fig3}b,c. In the ground state the Fermi level crosses only the hole and electron pockets and the system behaves as a topologically trivial metal. Indeed, a 3D metallic behaviour in WTe$_{2}$ has been reported to occur at very low temperatures (0.6K-4.2K)\cite{Zhu2015}. By raising temperature E$_F$ lifts to higher energies\cite{Wu2015,He2019,Wang2017,Pletikosic2014} and the hole pockets shrink and disappear at $\sim$40 meV  (Figures~\ref{Fig3}d-i).  Most important, at $\sim$40 meV the Fermi level crosses the Weyl bands in the vicinity of the nodal points. A pair of Weyl points (W1, W2), located along the K1-K2 line with k-space coordinates K1 = (0.1212, 0.0000, 0.0000) and K2 = (0.0929, 0.2000, 0.0000), is shown in  Figure~\ref{Fig3}c. At the same time arc-like FS segments are formed, connecting the electron pockets with the four symmetric pairs of Weyl nodal points, specified by white crosses in Figure~\ref{Fig3}g.  Evidently, the onset of the $\frac{1}{{T{_1}T}}$ $\sim$ T$^{2}$ temperature dependence at $\sim$120K marks the crossing of the Weyl bands and the transition from the metallic to the WF phase. 

\begin{figure}[h]
	\includegraphics[width=0.5\textwidth]{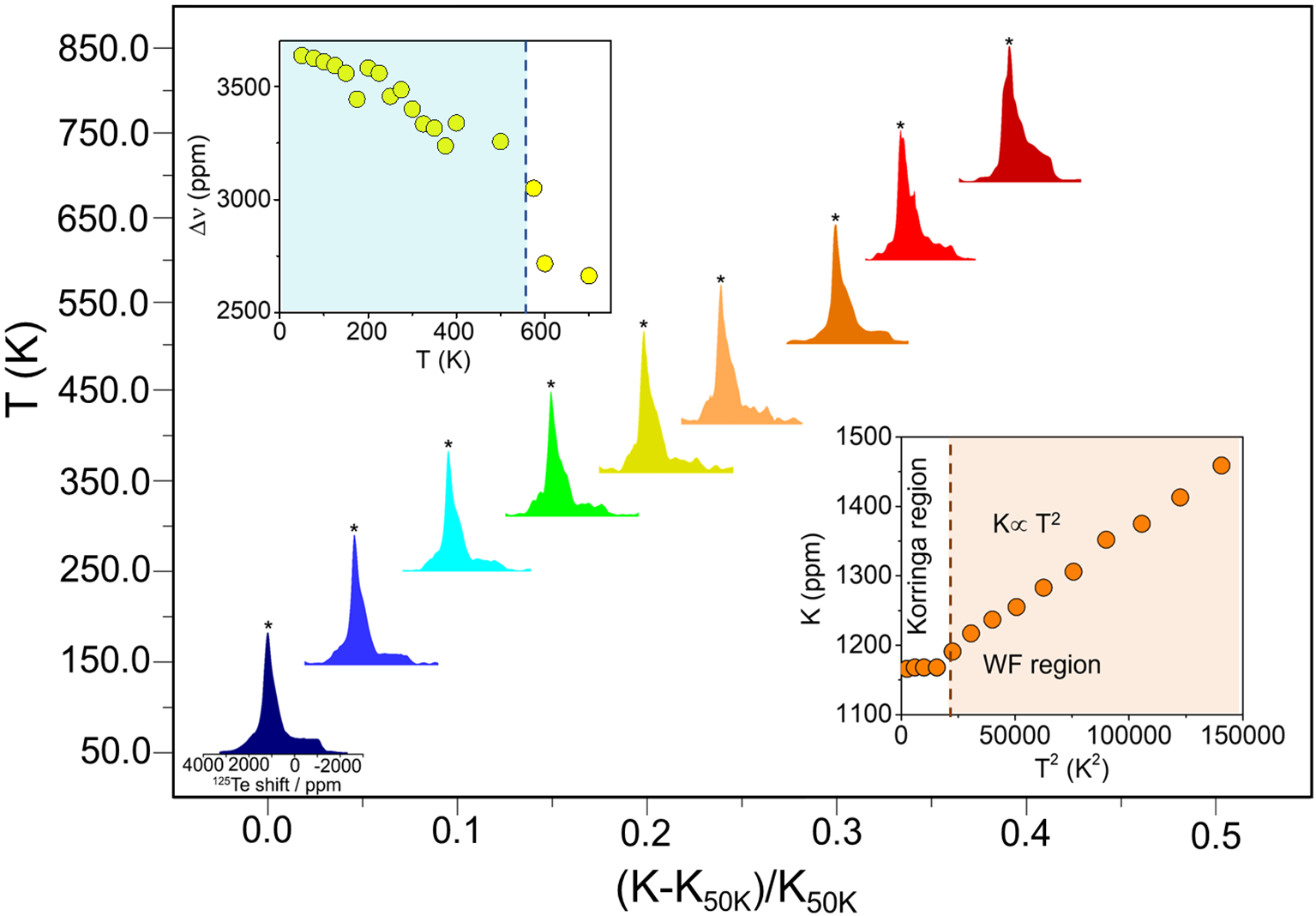}
	\caption{\label{Fig4} $^{125}$Te static NMR spectra in the temperature range 50K – 700K. Calibration in the frequency axis is performed according to the $K$-values of the most intense resonance as indicated by the star. The upper left inset shows the half height full width $\Delta\nu$ of the broad signal component assigned to the Te(1), Te(3) atomic sites.The lower right inset presents $K$ vs. $T^2$.}
	\end{figure}

To further enlighten the metal-to-WF phase transition, the  $K$ vs. $T$ trend was examined by measuring the $^{125}$Te static NMR spectra in the temperature range 50K – 700K. The main panel in Figure~\ref{Fig4} shows the frequency shift and the evolution of the lineshape upon heating, with spectra calibrated in the frequency axis according to the $K$ values of the most intense peak. Remarkably, $K$ exhibits the same temperature dependence as $\frac{1}{{T{_1}T}}$; At low temperatures it remains invariant, exhibiting Korringa metallic behaviour, while for T$>$120 it varies as $K\sim T^{2}$ (low right inset in  Figure~\ref{Fig4}), and subsequently shows a strong slope change at $\sim$ 400K. The quadratic temperature dependence is in contrast to both simple metals, where $K_{FC}$ dominates\cite{Papawassiliou2021,Korringa1950,Knight1949} and does not change with temperature, or Dirac and type-I Weyl semimetals with a point-like FS\cite{Xu2015,Lv2015}, where $K$ near the nodal points shows strong diamagnetic behaviour, varying according to formula\cite{Okvatovity2019} $K\sim-ln\frac{W}{k_{B}T}$, where W is a high energy cutoff.
A similar parabolic temperature dependence of $K$ has been observed in Dirac systems with tilted Dirac cones\cite{Hirata2017,Hirata2016}, that has been assigned to shrinkage of the FS and renormalization of $\upsilon$$_{F}$ by cooling. However, in the case of WTe$_{2}$, the linear Weyl bands coexist with the electron pockets and $K_{orb}$ prevails over $K_{FC}$, while in refs.\cite{Hirata2017,Hirata2016} only the isotropic K$_{FC}$ has been considered. Evidently, the simultaneous parabolic temperature dependence of $K$ and $\frac{1}{{T{_1}T}}$ in type-II WSMs is inherently related with the overtilted Weyl cones, which has been shown to be an extra source of relaxation enhancement\cite{Mohajerani2021}. 

In conclusion, high-resolution broadband NMR measurements on microcrystalline WTe$_{2}$ combined with DFT calculations illustrate excellently the topological band structure and the emergence of WFs. In this way an extraordinary Metal-to-WSM phase transition at  T$\approx$120K was revealed.  Undoubtedly, this kind of electron-states-resolved NMR crystallography appears to be very efficient in the study of low-energy quasiparticle excitations in type-II WSMs.  
\section{ACKNOWLEDGEMENTS}
W.P., J.P.C. and A.J.P. were supported by the Swedish Research Council (project no. 2016-03441) and the Swedish National Infrastructure for Computing (SNIC) through the center for parallel computing (PDC), project number 2019-3-500. 
\raggedright
\bibliography{WTe2_main}

\end{document}


\preprint{WTe2 NMR.TEX}  
\title{Detection of Weyl Fermions and the Metal to Weyl-Semimetal phase transition in WTe$_2$ via broadband High Resolution NMR}

\author{W. Papawassiliou}
\affiliation{Department of Materials and Environmental Chemistry, Arrhenius Laboratory, Stockholm University, Svante Arrhenius väg 16 C, SE-106 91 Stockholm, Sweden}
\author{J.P. Carvalho}
\affiliation{Department of Materials and Environmental Chemistry, Arrhenius Laboratory, Stockholm University, Svante Arrhenius väg 16 C, SE-106 91 Stockholm, Sweden}
\author{H.J. Kim}
\email[Corresponding author: ]{hansol@re.kbsi.kr}
\affiliation{Research Center for Materials Analysis, Korea Basic Science Institute, 169-148 Gwahak-ro, Yuseong-gu, Daejeon 34133, Republic of Korea}
\author{C.Y. Kim}
\affiliation{Research Center for Materials Analysis, Korea Basic Science Institute, 169-148 Gwahak-ro, Yuseong-gu, Daejeon 34133, Republic of Korea}
\author{S.J. Yoo}
\affiliation{Research Center for Materials Analysis, Korea Basic Science Institute, 169-148 Gwahak-ro, Yuseong-gu, Daejeon 34133, Republic of Korea}
\author{J.B. Lee}
\affiliation{Research Center for Materials Analysis, Korea Basic Science Institute, 169-148 Gwahak-ro, Yuseong-gu, Daejeon 34133, Republic of Korea}
\author{S. Alhassan}
\affiliation{Department of Chemical Engineering, Khalifa University, PO Box 2533, Abu Dhabi, United Arab Emirates}
\author{S. Orfanidis}
\affiliation{Institute of Nanoscience and Nanotechnology, NCSR Demokritos, 15310 Aghia Paraskevi, Attiki, Greece}
\author{V.Psycharis}
\affiliation{Institute of Nanoscience and Nanotechnology, NCSR Demokritos, 15310 Aghia Paraskevi, Attiki, Greece}
\author{M. Karagianni}
\affiliation{Institute of Nanoscience and Nanotechnology, NCSR Demokritos, 15310 Aghia Paraskevi, Attiki, Greece}
\author{M. Fardis}
\affiliation{Institute of Nanoscience and Nanotechnology, NCSR Demokritos, 15310 Aghia Paraskevi, Attiki, Greece}
\author{N. Panopoulos}
\affiliation{Institute of Nanoscience and Nanotechnology, NCSR Demokritos, 15310 Aghia Paraskevi, Attiki, Greece}
\author{G. Papavassiliou}
\email[Corresponding author: ]{g.papavassiliou@inn.demokritos.gr}
\affiliation{Institute of Nanoscience and Nanotechnology, NCSR Demokritos, 15310 Aghia Paraskevi, Attiki, Greece}
\author{A.J. Pell}
\email[Corresponding author: ]{andrew.pell@ens-lyon.fr}
\affiliation{Centre de RMN à Très Hauts Champs de Lyon (UMR 5280 CNRS/ENS Lyon/Université Claude Bernard Lyon 1), Université de Lyon, 5 rue de la Doua, 69100 Villeurbanne, France}

\pacs{73.20.-r,76.60.Cq, 82.56.Fk, 82.56.Na}
\keywords{topological matter, type-II Weyl Semimetals, solid-state NMR, Lifshitz transition}

\date{\today}

\maketitle

In this supplementary information, we provide additional details on certain aspects of the study reported in the main manuscript.\\
\indent 
The following issues are discussed:\\
\indent  
1. Materials and Methods.\\
\indent 
2. XRD and TEM sample characterization.\\
\indent 
3. DFT calculated energy bands of the T$_{d}$ and 1T$^\prime$ crystal phases.\\
\indent 
4.  $^{125}$Te ssNMR measurements and DFT calculations of the Knight shift. \\
\indent
5. The evolution of the Te projected Density of States (pDOS) close to the Fermi energy. \\
\section{1. MATERIALS AND METHODS}
\subsection{Sample and XRD characterization} The microcrystalline WTe$_2$ sample was purchased from the TRUNNANO, Luoyang Tongrun Nano Technology Co. The sample quality was checked with XRD and TEM analysis. Powder x-ray diffraction patterns were collected at room temperature using an x-ray diffractometer (XRD, Malvern Panalytical B.V., Empyrean) operated at 40 kV and 30 mA of $CuK\alpha$ radiation ($\lambda$ = 1.5406 \AA) in 5–95$^\circ$ 2$\theta$ range using a step size of 0.013$^\circ$. Rietveld refinement\cite{Rietveld1969} was performed using the HighScore Plus (Malvern Panalytical B.V.) software package. 
\subsection{TEM experiments} To verify the structural phase transition of WTe$_2$, in situ heating transmission electron microscopy (TEM) study was performed. A cross sectional TEM specimen with a specific direction of $<010>$ was firstly prepared from a rectangular shape (6.5$\mu$m × 5.0$\mu$m) of WTe$_2$ sample using focused ion beam system (FIB, Quanta 3D FEG, FEI). A high voltage electron microscope (HVEM, Jeol Ltd., JEM ARM 1300S) with a double tilt heating holder (Gatan Inc., 652) was used for the in situ heating experiment. The temperature was increased up to 500, 550 and 600 K with a heating rate of 10 K/min and maintained for about 20 minutes before acquiring high resolution TEM (HRTEM) images to minimize specimen drift. Lattice parameter and angle between crystal planes were measured from fast Fourier transform (FFT) patterns of the individual HRTEM images. The simulated electron diffraction pattern (EDP) and atomic model were built by CrystalMaker program (CrystalMaker Software Ltd.).
\subsection{NMR experiments} The $^{125}$Te MAS experiments on microcrystalline WTe$_2$ were performed at three magnetic field strengths on three different spectrometers. The static NMR experiments were performed in magnetic field 9.4 T.

(i) \underline{MAS NMR experiments at 9.4 T}. The $^{125}$Te MAS spectra of WTe$_2$ were collected with a 2.5 mm HX double-resonance probe, at 30 kHz MAS on a Bruker 400 Avance-III spectrometer operating at a $^{125}$Te Larmor frequency of 126.23 MHz. Spectral acquisition was done with a double adiabatic spin-echo sequence (DAE) with a 1.6 $\mu$s 90$^\circ$ excitation pulse length, proportional to an RF field of 156 kHz, followed by a pair of rotor-synchronized short high-power adiabatic pulses (SHAPs)\cite{Carvalho2021,Pell2019} of 33.33 $\mu$s length and a 5 MHz frequency sweep. For the separation of the isotropic shift and chemical shift anisotropy, which is a matter of significance in our case and in heavy spin-1/2 nuclei in general, the adiabatic magic-angle-turning (aMAT) pulse sequence was employed\cite{Clement2012}. The aMAT consists of a $\pi$/2 excitation pulse followed by 6 refocusing SHAP $\pi$-pulses. The same SHAPs as for the double adiabatic spin-echo sequence were used. Separation of isotropic Knight Shifts was achieved in the isotropic “infinite speed” dimension, whereas the MAS dimension corresponds to the conventional MAS spectrum. The constant evolution time of the aMAT experiment was 66.66 $\mu$s, excluding the length of the SHAPs, which is equivalent to two rotor periods. The recycle delay for all experiments was set to 80 ms, in accordance with 5$\cdot$T$_1$.

(ii) \underline{ MAS NMR experiments at 14.1 T}. The  $^{125}$Te MAS spectra of WTe$_{2}$ were acquired with a 1.3 mm HX probe, at 60 kHz and 30 kHz MAS, on a MAS on a Bruker 600 Avance-III spectrometer operating at a  $^{125}$Te Larmor frequency of 189.339 MHz. In the case of 60 kHz MAS, for the acquisition of the DAE spectrum, an initial excitation pulse with length 1.1 $\mu$s corresponding to RF field amplitude of 227 kHz was used, and the following rotor synchronized SHAPs were sweeping through 5 MHz in 33.33 $\mu$s. For the aMAT spectra, the same SHAPs were used combined with an evolution time of 66.66 $\mu$s, excluding the length of the SHAPs, which is equivalent to four rotor periods. For 30 kHz MAS, the same SHAPs and excitation pulse was used for both for the DAE and the aMAT experiments, with the only the evolution time changing in the case of the aMAT experiment, i.e. 66.66 $\mu$s corresponding to 2 rotor periods. For all experiments a recycle delay of 80 ms was used.

(iii)  \underline{MAS NMR experiments at 23.5 T}. The  $^{125}$Te MAS spectra of WTe$_2$ were collected with a 2.5 mm HX probe, at 30 kHz on a Bruker 1000 Avance Neo spectrometer operating at a  $^{125}$Te Larmor frequency of 315.570 MHz. For the acquisition of all spectra the excitation pulse was of length 1.75 $\mu$s, corresponding to an RF field of 142.8 kHz. The following SHAPs swept through 5 MHz in 33.33 $\mu$s. In all cases, chemical shifts were referenced to TeO$_2$37. 

(iv)  \underline{Static NMR experiments at variable temperature}.  The variable temperature, frequency-sweep  $^{125}$Te NMR spectra were acquired on a home-built NMR spectrometer under static conditions, operating at Larmor frequency of 126.23 MHz. An Oxford 1200CF continuous flow cryostat was employed for measurements in the temperature range 50-400K and an Oxford HT1000V furnace for measurements in the range 400K-700K. For the spin-lattice relaxation time T$_1$ experiments a $\pi$/2-t-$\pi$/2-$\tau$-$\pi$ saturation recovery pulse sequence was implemented.

\subsection{DFT Calculations} The Quantum Espresso package\cite{Giannozzi2009} was used to carry out DFT calculations of the k-resolved projected density of states of the orbitals for bulk WTe$_2$. Calculations were performed on the basis of the Perdew-Burke-Ernzerhof (PBE) type generalized gradient approximation\cite{Perdew1996}. For the Brillouin zone integrations we used a 11x11x1 Monkhorst-Pack k-point mesh\cite{Monkhorst1976}, and the kinetic energy cutoff was fixed to 800 eV. The lattice constants were acquired by the Rietveld refinement of the XRDs (a=6.282 \AA,  b= 3.469 \AA, c= 14.07 \AA). Spin-orbit effects were treated self-consistently using fully relativistic Projector Augmented Wave (PAW) pseudopotentials\cite{Bloechl1994}. The Fermi Surface analysis was performed with the FermiSurfer software package, with data files generated by Quantum Espresso.  

Energy bands structure and NMR Knight Shift calculations were performed by using the full-potential linearized augmented plane-wave method, as implemented in the Wien2k DFT software package\cite{Blaha2001}. The spin-orbit interaction was considered in a second variational method. Calculations were performed with and without spin-orbit coupling on bulk WTe$_{2}$. The k-mesh convergence was checked up to 100,000 points. Other computational parameters like atomic sphere radii as well as potentials and wave functions inside the atomic spheres are kept as set by Wien2k defaults. The plane wave basis set size was determined by seting RKmax=8, and for presented results we have used the PBE generalized gradient approximation. The orbital part of the Knight shift K$_{orb}$ was calculated by using the x nmr script of the Wien2k software package, by activating switches to include SOC, and Fermi-Dirac smearing between 2-8 mRy. The Fermi Contact and dipolar terms were calculated in the presence of SOC, using a spin-polarized set-up as explained in ref.\cite{Blaha2001}.

\subsection{Simulation of the aMAT NMR spectrum} Numeric simulations of the aMAT NMR spectra were performed in SIMPSON\cite{Bak2000}, using the Zaremba, Conroy, Wolfsberg and co-workers (ZCW) powder averaging scheme\cite{Cheng1973} with 615 crystallites, 40 gamma angles and employing Gaussian line broadening. The 2D spectrum was simulated considering MAT\cite{Hu1993}, by using ideal $\pi$ pulses. Each one of the four Te environments was simulated using a Gaussian distribution of isotropic shifts, with a standard deviation of 15.7 ppm, centered at the DFT predicted isotropic shift positions.
\section{2. XRD and TEM sample characterization}
The XRD pattern of the microcrystalline WTe$_2$ sample at room temperature is shown in Supplementary Figure~\ref{FigS1}. Rietveld refinement \cite{Rietveld1969} was performed with the help of the HighScore Plus (Malvern Panalytical B.V.) software package to determine the purity of the sample. The analysis indicates that the sample consists primarily of $\gamma$-WTe$_2$ (orthorhombic phase T$_{d}$, space group Pmn2$_{1}$, \#31, with cell dimensions a = b = 5.858(2) {\AA} and c = 3.384(5) \AA) \cite{Dawson1987} with traces of elemental cubic W (Im$\overline{3}$m, \#93 with cell dimensions a = b = c = 2.760 \AA) and elemental Te (trigonal P3$_{1}$2$_{1}$, \#152 with cell dimensions a = b = 4.512 {\AA}  and c = 5.960 \AA).
For further experimental verification of the structural phase transition from the orthorhombic T$_{d}$ phase (space group \#31 , Pmn2$_{1}$) to the monoclinic 1T$^\prime$ (space group \#11, P2$_{1}$/m)\cite{Tao2020} of microcrystalline WTe$_{2}$, in situ HRTEM measurements were performed. For the measurements, a cross sectional TEM specimen with a specific direction of [010] was firstly prepared from a rectangular (6.5 $\mu$m × 5.0 $\mu$m) WTe$_{2}$ grain using a focused ion beam system (FIB, Quanta 3D FEG, FEI), as shown in Supplementary Figure~\ref{FigS2}. The [010] direction was chosen because the primary difference between the two space groups lies on the fact that the $\beta$ angle in the monoclinic space group deviates from the 90$^\circ$ in comparison to the orthorhombic Pmn2$_{1}$.
\begin{figure}[h]
	\includegraphics[width=0.8\textwidth]{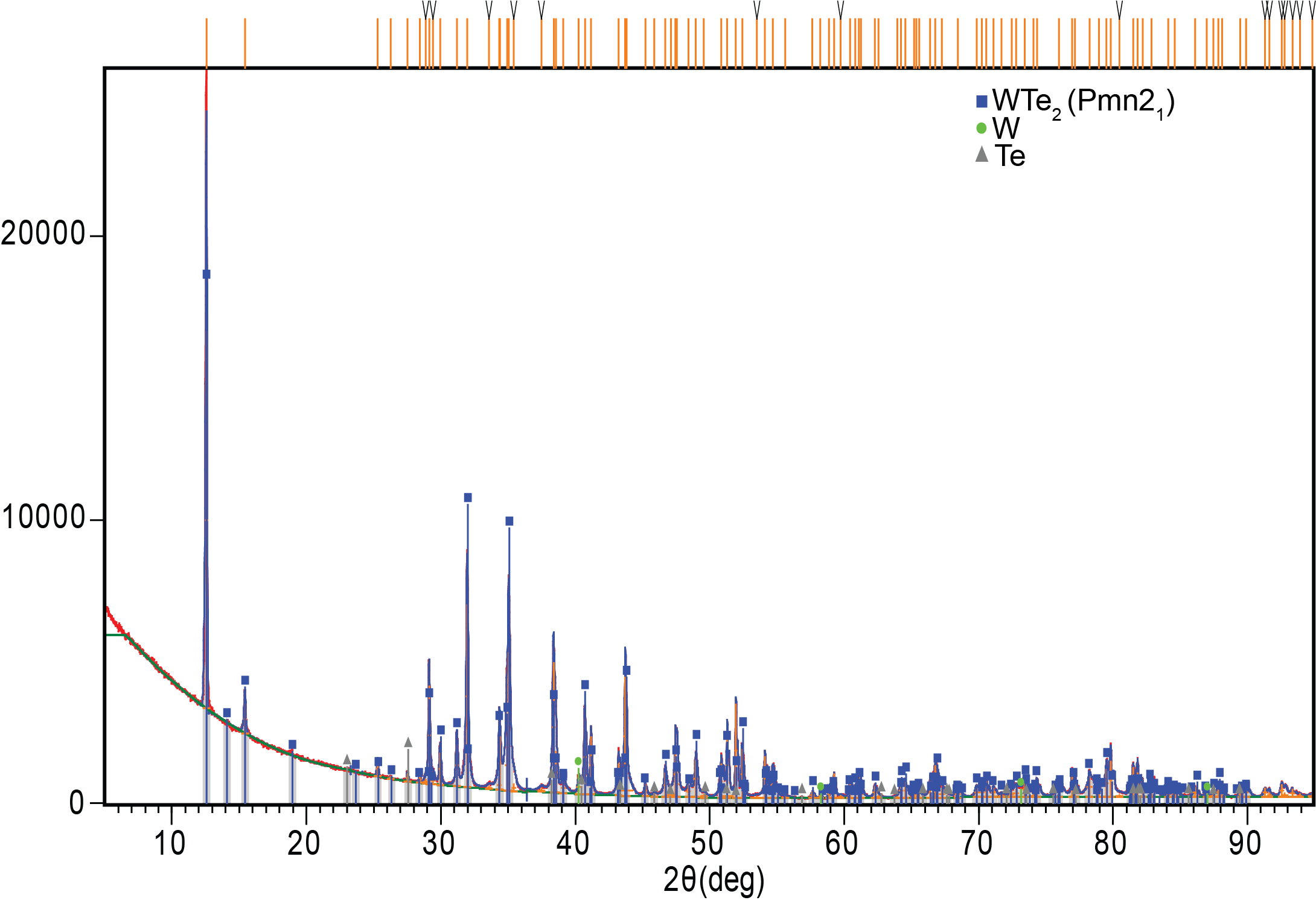}
	\caption{\label{FigS1}The experimental X–ray powder diffraction patterns (red) and the Rietveld analysis (blue) of microcrystalline WTe$_{2}$ crystallized in the orthorhombic Pmn2$_{1}$ space group (S.G. \#31). Traces of elemental W (green) and Te (gray) are also observed.}
	\end{figure}
\indent
\begin{figure}[h]
	\includegraphics[width=0.8\textwidth]{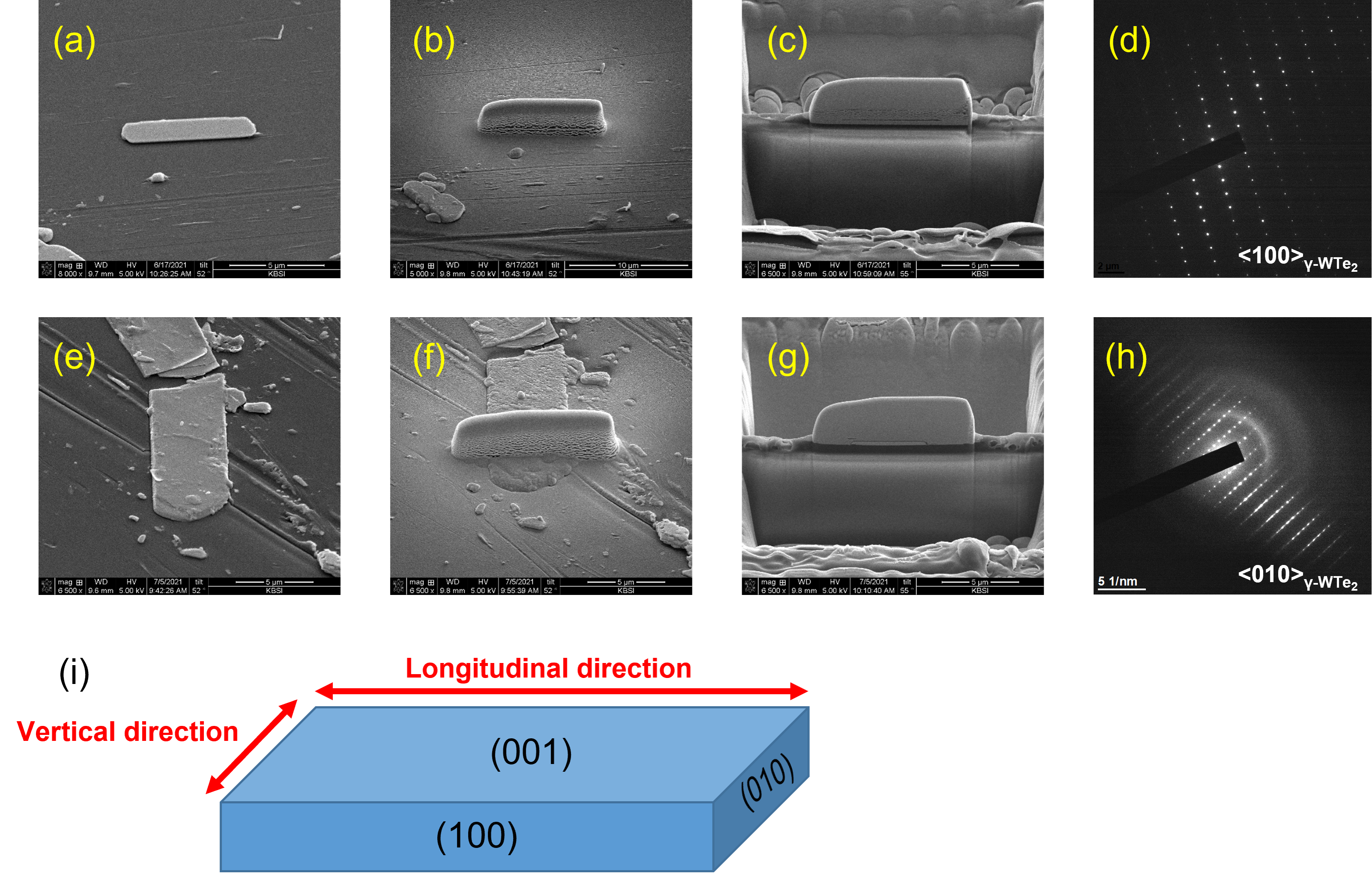}
	\caption{\label{FigS2} Sample preparation for the in-situ heating HRTEM by employing the focused ion beam (FIB) method of microcrystalline WTe$_{2}$:(a-c) Calibration of the sample along the longitudinal (100) plane with the aid of FIB and (d) the corresponding electron diffraction pattern along the [100] direction. (e-g) Calibration of the sample along the vertical (010) plane in a similar manner as mentioned above and (h) the corresponding electron diffraction pattern. (i) Depiction of the platelet-like flake on which FIB is performed, with labelled facets.}
	\end{figure}
\indent
\begin{figure}[h]
	\includegraphics[width=0.8\textwidth]{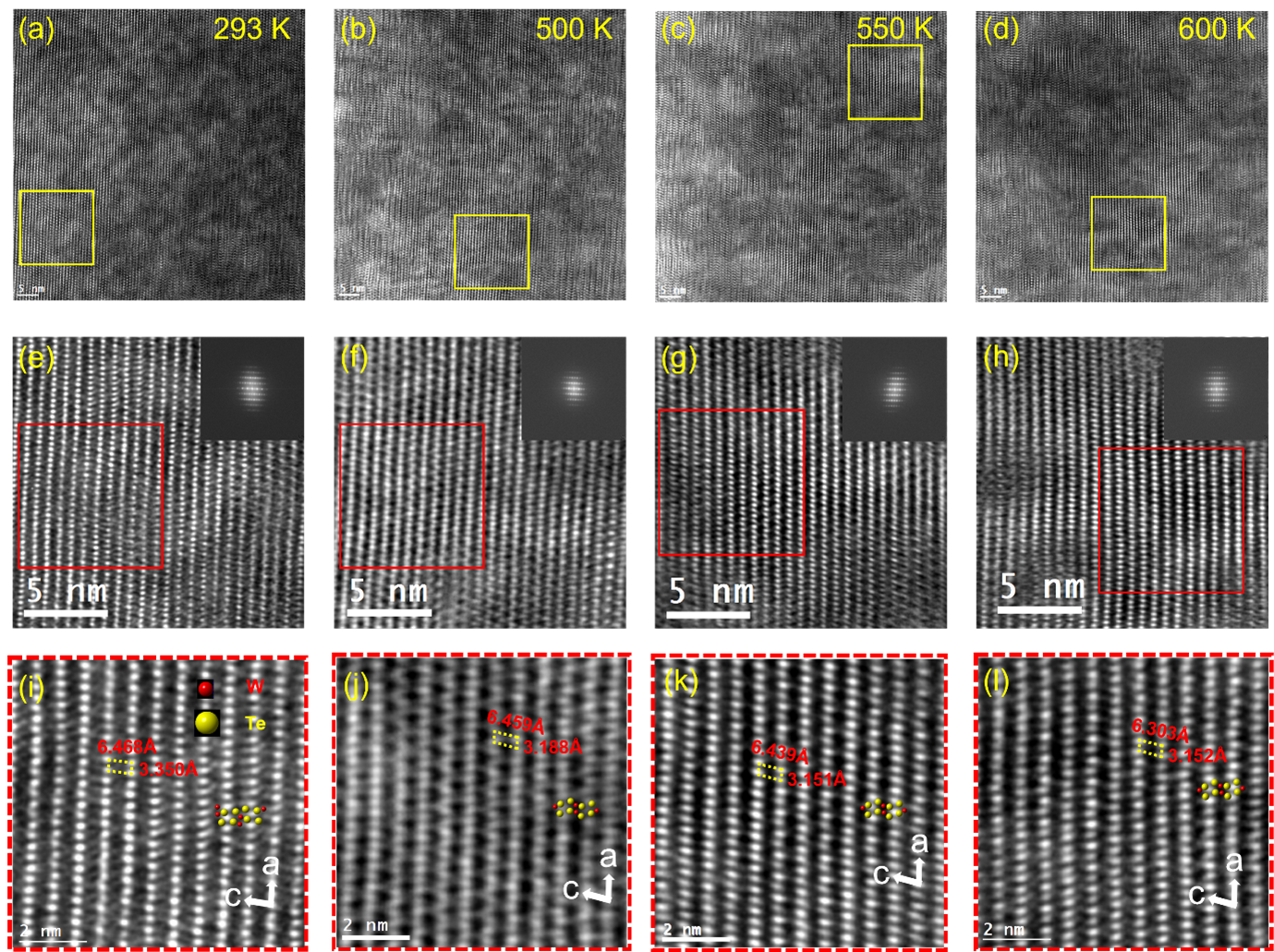}
	\caption{\label{FigS3} HRTEM analysis of a plate-shaped particle in the microcrystalline WTe$_{2}$ sample from room temperature (293K) up to 600K. (a-d) HRTEM of microcrystalline WTe$_{2}$ along the [010] direction, at 293K, 500K, 550K, 600K. (e-h) Filtered Image and FFT data along the [010] direction for the same temperatures. The red rectangle highlights the areas expansions of which are presented in (i-l) to showcase the change in the crystal lattice parameters upon heating.}
	\end{figure}
\indent
\begin{figure}[h]
	\includegraphics[width=0.8\textwidth]{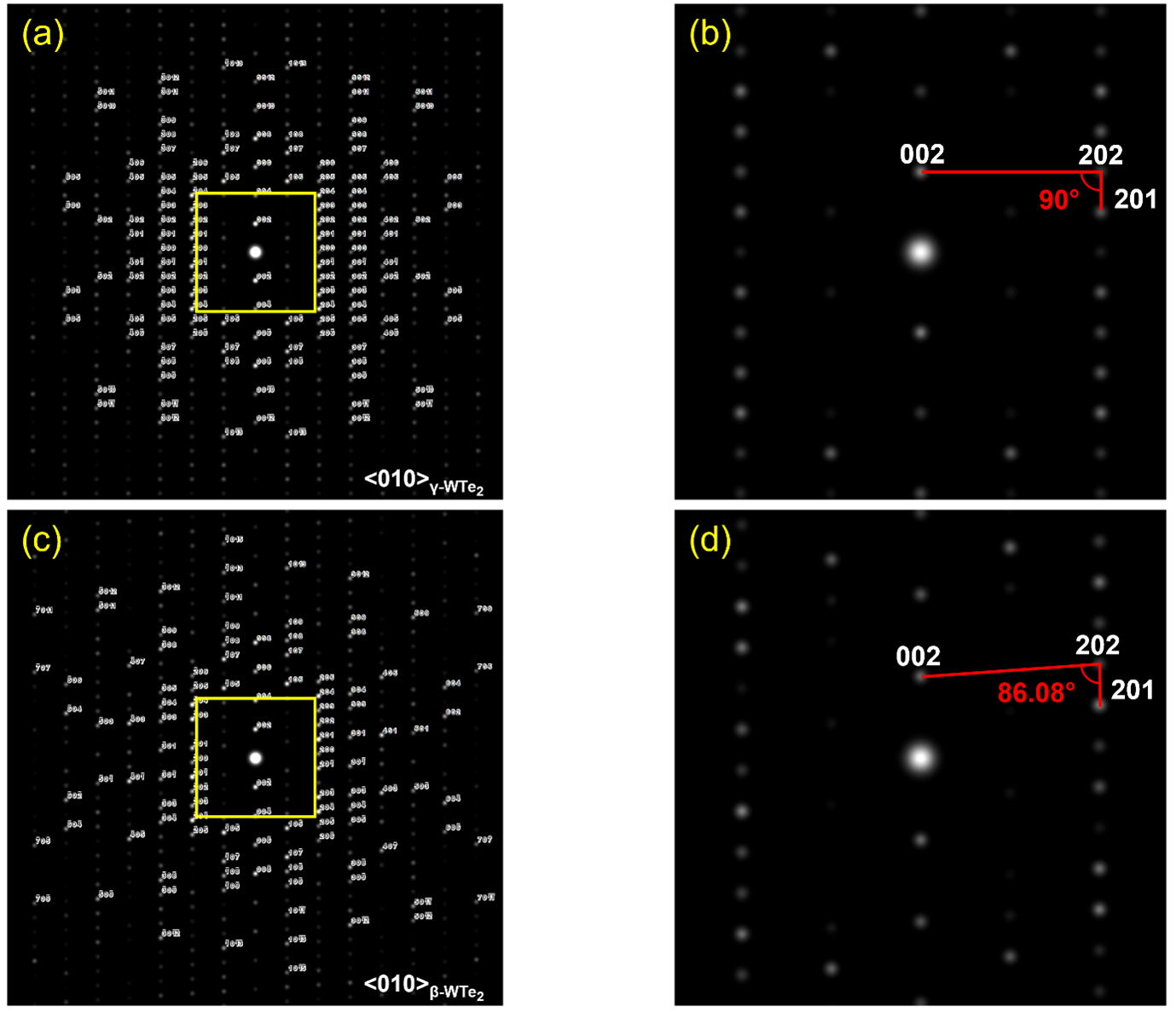}
	\caption{\label{FigS4}(a-d) Simulated Electron Diffraction Patterns for the orthorhombic $\gamma$-WTe$_{2}$ and the monoclinic $\beta$-WTe$_{2}$. (a) Simulated EDP of WTe$_{2}$ in the orthorhombic Pmn2$_{1}$ phase along the $<010>$ zone axis. (b) Simulated EDP of WTe$_{2}$ in the monoclinic P2$_{1}$/m phase along the $<010>$ zone axis. (c,d) Expansions of the EDP highlighting the diffraction peaks of the 002, 202 and 201 reflections along the $<010>$ zone axis. }
	\end{figure}
Upon heating from RT to initially 500 K, a temperature prior to the phase transition, a contraction in the lattice constant a from 3.350 {\AA}  to 3.188 {\AA}   is the only noticeable difference in the filtered HRTEM images presented in Supplementary Figures~\ref{FigS3}i,j. In the temperature range 550 K to 600 K, the lattice constant c also contracts from 6.439 {\AA}  to  6.303 {\AA}   (Supplementary Figure~\ref{FigS3}k,l).Although these changes in the lattice constants are strong indications of structural changes, further definitive evidence is needed for the phase transition. 
\indent
\begin{figure}[h]
	\includegraphics[width=0.8\textwidth]{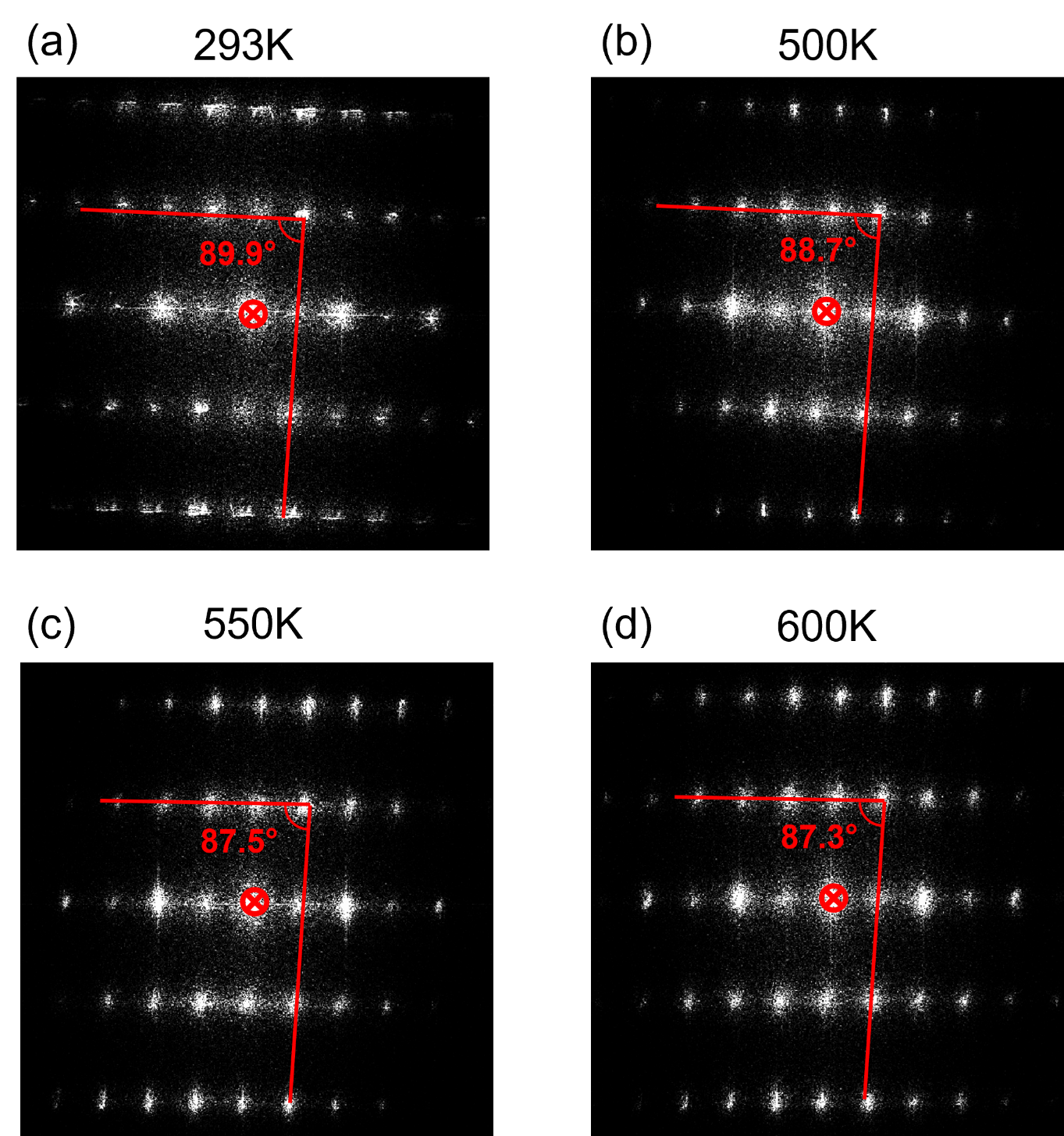}
	\caption{\label{FigS5}Experimental electron diffraction patterns along the $<010>$ zone axis highlighting the structural phase transition from $\gamma$-WTe$_{2}$ to $\beta$-WTe$_{2}$. (a-d) Experimental EDPs of WTe$_{2}$ at temperatures of (a) 293 K, (b) 500 K, (c) 550 K, and (d) 600 K.  }
	\end{figure}
From the simulated electron diffraction patterns along the $<010>$ zone axis it becomes evident that upon phase transition, the 002 peak is displaced from an 90$^\circ$ alignment with the 202 and 201 peaks, as is characteristic for orthorhombic space groups (c), to a lower angle, 86.08$^\circ$ indicative of the sample entering a monoclinic P2$_{1}$/m space group. The simulated EDP’s, presented in Supplementary Figure~\ref{FigS4}, corroborate the experimental findings in Supplementary Figure~\ref{FigS5}, where the phase transition is observable by the change of the $\beta$ angle from 89.9$^\circ$ for the orthorhombic phase Pmn2$_{1}$ at 293 K to 87.3$^\circ$ for the monoclinic phase P2$_{1}$/m at 600 K.
\section{3. DFT calculated energy bands of the T$_{d}$ and 1T$^\prime$ crystal phases.}
\indent
\begin{figure}[h]
	\includegraphics[width=0.8\textwidth]{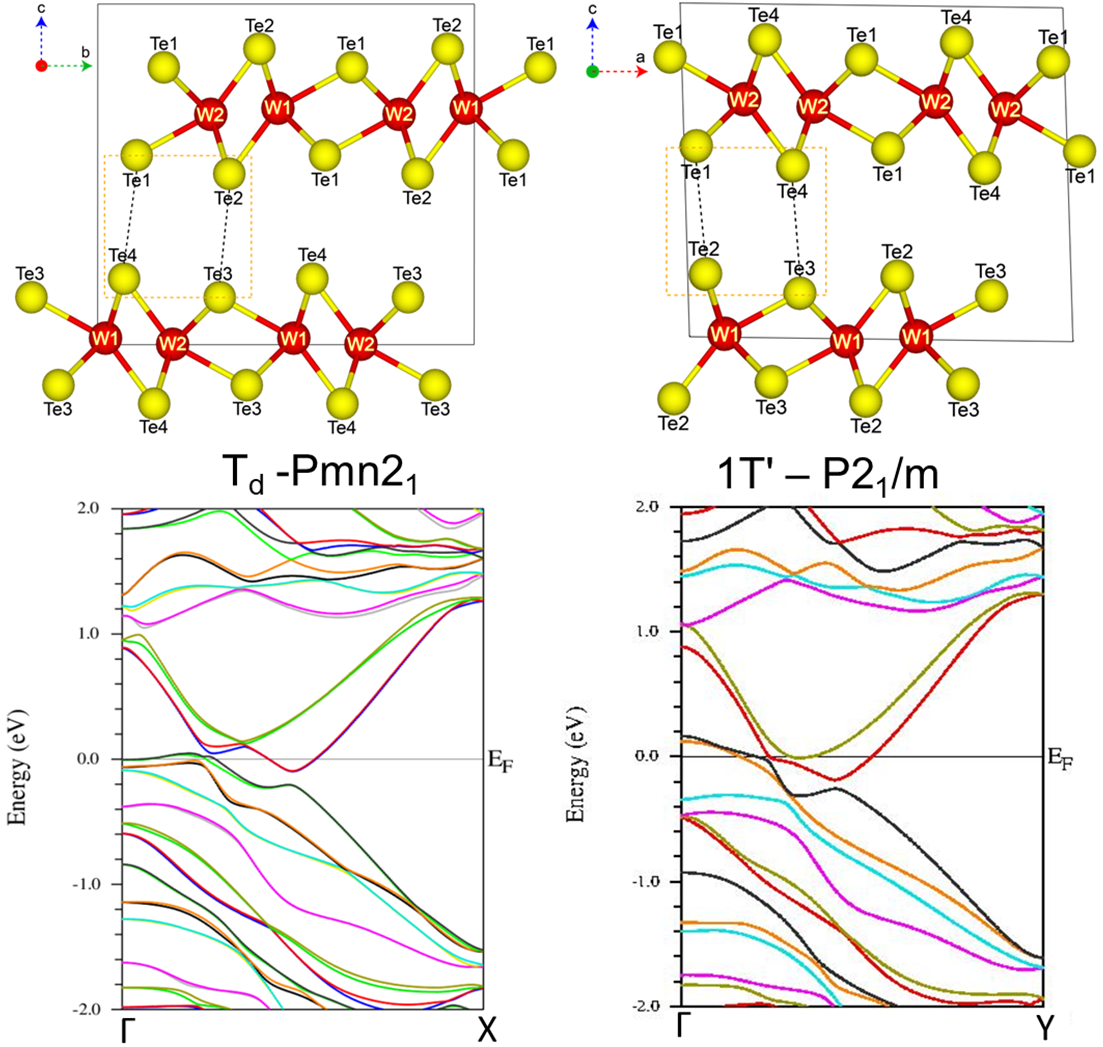}
	\caption{\label{FigS6}The unit cells and the electron energy band structures of the WTe$_{2}$ T$_{d}$-Pmn2$_{1}$ and 1T$^\prime$-P2$_{1}$/m crystal phases, along the lines connecting the high symmetry points $\Gamma$$\rightarrow$X and  $\Gamma$$\rightarrow$Y, respectively.}
	\end{figure}
Supplementary Figure~\ref{FigS6} compares the band structure of the T$_{d}$ and 1T$^\prime$ crystal phases along the $\Gamma$ $\rightarrow$ X and $\Gamma$ $\rightarrow$ Y high symmetry lines. Despite the minor changes in the crystal structure, the band structures of the 1T$^\prime$ crystal phase near the Fermi level changes significantly in comparison to the low temperature T$_d$ crystal phase. 
\section{4. $^{125}$T$e$ $ss$NMR measurements and DFT calculations of the Knight shift. }
$^{125}$Te ssNMR experiments were performed under both magic-angle spinning (MAS) at 30 kHz and static conditions as shown in Supplementary Figure~\ref{FigS7}, revealing an anisotropic lineshape spanning approximately 3000 ppm, and featuring a maximum at 1200 ppm. The MAS spectrum shows  a splitting at this maximum, due to the small differences in the local environments of the almost-crystallographically-equivalent pairs Te(1), Te(3) and Te(2), Te(4). However, both the precise isotropic Knight shifts and the relevant shift anisotropies are still to be determined. We subsequently performed DFT calculations of the NMR Knight shifts with the aid of the WIEN2k package\cite{Blaha2001}, as presented in Supplementary Table 1. 
\begin{table}[h]
\begin{tabular}{|
>{\columncolor[HTML]{34CDF9}}l 
>{\columncolor[HTML]{ECF4FF}}c 
>{\columncolor[HTML]{ECF4FF}}c 
>{\columncolor[HTML]{ECF4FF}}c 
>{\columncolor[HTML]{ECF4FF}}c 
>{\columncolor[HTML]{ECF4FF}}c 
>{\columncolor[HTML]{ECF4FF}}c 
>{\columncolor[HTML]{ECF4FF}}l |}
\hline
\multicolumn{4}{|l}{\cellcolor[HTML]{34CDF9}Calculated NMR shift}                                                                                                                       & \multicolumn{4}{c|}{\cellcolor[HTML]{34CDF9}Experimental NMR shift}                                                                \\ \hline
\multicolumn{1}{|l|}{\cellcolor[HTML]{34CDF9}\textbf{Site}}  & \cellcolor[HTML]{68CBD0}$K_{iso}$ (calc)/ppm & \cellcolor[HTML]{68CBD0}$\Delta$(calc) / ppm  & \cellcolor[HTML]{68CBD0}$\eta$             & \cellcolor[HTML]{68CBD0}$K_{iso}$ (exp)/ppm & \cellcolor[HTML]{68CBD0}$\Delta$(exp) / ppm & \multicolumn{2}{c|}{\cellcolor[HTML]{68CBD0}$\eta$}     \\ \hline
\multicolumn{1}{|l|}{\cellcolor[HTML]{34CDF9}\textbf{Te(1)}} & 1560                                    & -2716                                  & 0.923                                 & 1080                                   & -2305.8                              & \multicolumn{2}{c|}{\cellcolor[HTML]{ECF4FF}0.53}  \\
\multicolumn{1}{|l|}{\cellcolor[HTML]{34CDF9}\textbf{Te(2)}} & 1750                                    & -908                                   & 0.817                                 & 1269                                   & -855.51                              & \multicolumn{2}{c|}{\cellcolor[HTML]{ECF4FF}0.117} \\
\multicolumn{1}{|l|}{\cellcolor[HTML]{34CDF9}\textbf{Te(3)}} & 1608                                    & -2574                                  & 0.851                                 & 1128                                   & -2861.5                              & \multicolumn{2}{c|}{\cellcolor[HTML]{ECF4FF}0.5}   \\
\multicolumn{1}{|l|}{\cellcolor[HTML]{34CDF9}\textbf{Te(4)}} & 1697                                    & -1100                                  & 0.252                                 & 1351                                   & -786.3                               & \multicolumn{2}{c|}{\cellcolor[HTML]{ECF4FF}0.398} \\ \hline
\multicolumn{1}{|l|}{\cellcolor[HTML]{34CDF9}\textbf{}}      & \multicolumn{7}{c|}{\cellcolor[HTML]{34CDF9}Calculated NMR shielding}                                                                                                                                                                                         \\ \hline
\multicolumn{1}{|l|}{\cellcolor[HTML]{34CDF9}\textbf{Site}}  & \cellcolor[HTML]{68CBD0}$\sigma_{orb}$ (calc)/ppm & \cellcolor[HTML]{68CBD0}$\sigma_{FC}$ (calc)/ppm & \cellcolor[HTML]{68CBD0}Orbital - $\sigma_{xx}$ & \cellcolor[HTML]{68CBD0}Orbital - $\sigma_{yy}$  & \multicolumn{3}{c|}{\cellcolor[HTML]{68CBD0}Orbital - $\sigma_{zz}$}                                \\ \hline
\multicolumn{1}{|l|}{\cellcolor[HTML]{34CDF9}\textbf{Te(1)}} & 778                                     & 32                                     & -963.1                                & 708.83                                 & \multicolumn{3}{c|}{\cellcolor[HTML]{ECF4FF}2589.07}                                      \\
\multicolumn{1}{|l|}{\cellcolor[HTML]{34CDF9}\textbf{Te(2)}} & 1510                                    & -890                                   & 960.81                                & 1455.54                                & \multicolumn{3}{c|}{\cellcolor[HTML]{ECF4FF}2116.63}                                      \\
\multicolumn{1}{|l|}{\cellcolor[HTML]{34CDF9}\textbf{Te(3)}} & 698                                     & 64                                     & -890.52                               & 570.97                                 & \multicolumn{3}{c|}{\cellcolor[HTML]{ECF4FF}2415.14}                                      \\
\multicolumn{1}{|l|}{\cellcolor[HTML]{34CDF9}\textbf{Te(4)}} & 1518                                    & -845                                   & 1059.65                               & 1244.26                                & \multicolumn{3}{c|}{\cellcolor[HTML]{ECF4FF}2252.03}                                      \\ \hline
\end{tabular}
\caption{Comparison of calculated and experimental $^{125}$Te NMR parameters of the microcrystalline WTe$_2$ sample. $K_{calc}=\sigma_{ref}-\sigma_{calc}$ with $\sigma_{ref}$=2370.1 ppm as derived in ref.\cite{Papawassiliou2020}, and $\sigma_{calc} = \sigma_{FC}+ \sigma_{dip}+ \sigma_{orb}$. The dipolar contribution is negligible. Therefore, only the principal values of the orbital shielding tensor $\sigma_{xx}$, $\sigma_{yy}$, and $\sigma_{zz}$ are presented as it is the only source of anisotropy. Calculated shift anisotropies are derived from the Haeberlen convention\cite{1976}, where $\Delta = \sigma_{zz} – (\sigma_{xx}+\sigma_{yy})/2$, and asymmetry parameter $\eta=(\sigma_{yy}-\sigma_{xx})/(\sigma_{zz}-\sigma_{iso})$.}
\label{tab:my-table}
\end{table}
 Results show that Te(1),Te(3) exhibit a broad anisotropic pattern, whereas Te(2),Te(4) are more shifted towards positive frequencies. Overall, a very good match between calculated and experimental spectra is observed. The origin of the anisotropy stems from the almost pure Weyl $\ket{p}$ orbitals of the Te(1), Te(3) sites, whereas Te(2) and Te(4) show increased Fermi Contact shielding. Calculated Knight shift values are given by expression $K_{calc}$=$\sigma_{ref}$ - $\sigma_{calc}$ with $\sigma_{calc}$ = $\sigma_{FC}$ + $\sigma_{dip}$ + $\sigma_{orb}$ and $\sigma_{ref}$ =2370.1 ppm, as derived in ref.\cite{Papawassiliou2020}. Calculated shift anisotropies are presented in the Haeberlen convention\cite{1976}.  The dipolar contribution was found to be negligibly small. Furthermore, the influence of the magnetic field strenght on the NMR spectra was examined. Supplementary Figure~\ref{FigS8} shows the 1-D $^{125}$Te MAS ssNMR experimental and DFT-calculated spectra at 9.4 T, 14.1 T and 23.5 T.  
\begin{figure}[h!]
	\includegraphics[width=0.7\textwidth]{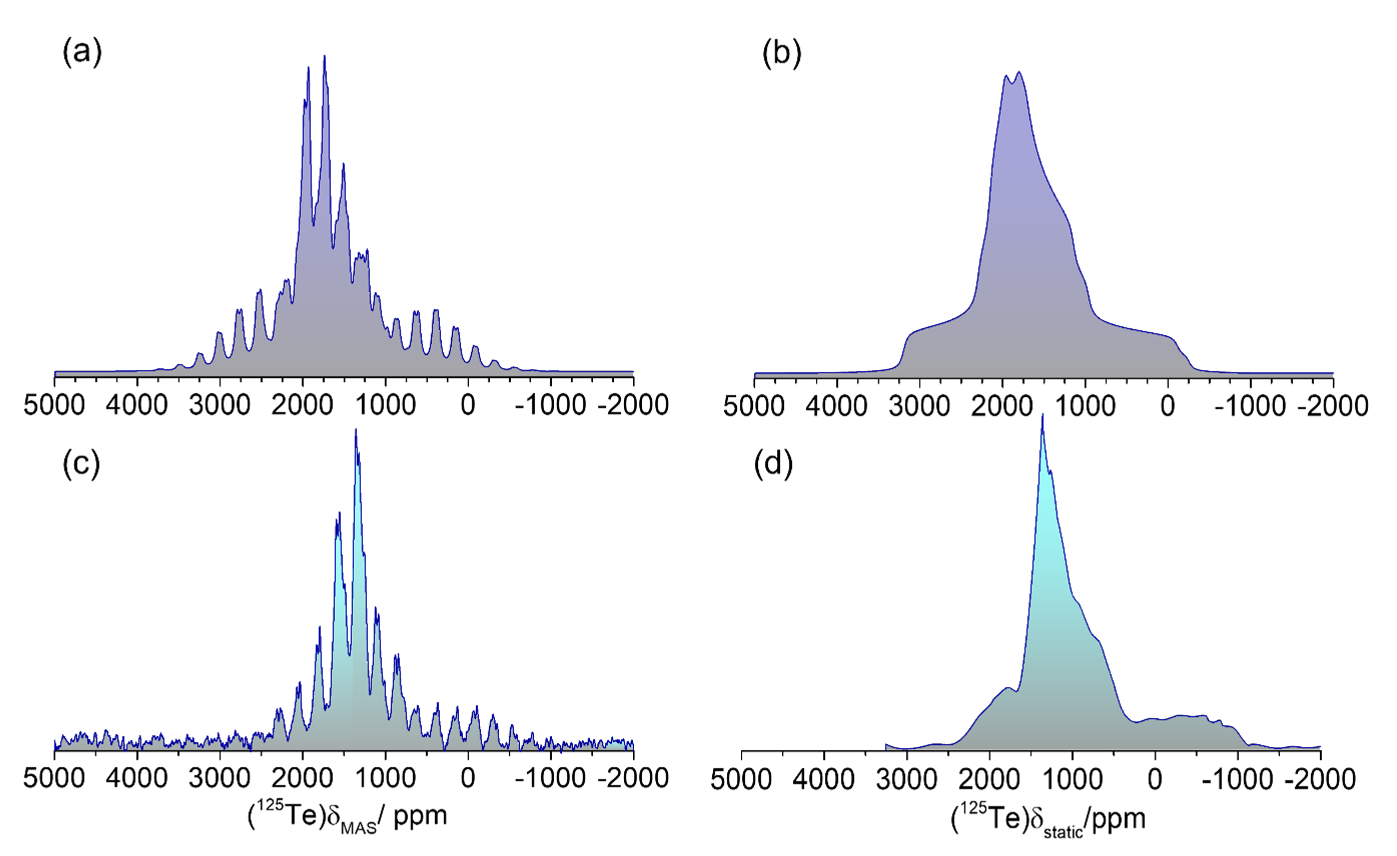}
	\caption{\label{FigS7}DFT calculated and experimental solid-state NMR spectra of microcrystalline WTe$_{2}$ at 9.4T. (a,b) The DFT calculated ssNMR spectra under magic angle spinning (a) and static (b) conditions. (c,d) The experimental ssNMR spectra under magic angle spinning (c) and static (d) conditions.}
	\end{figure}
\begin{figure}[h!]
	\includegraphics[width=0.7\textwidth]{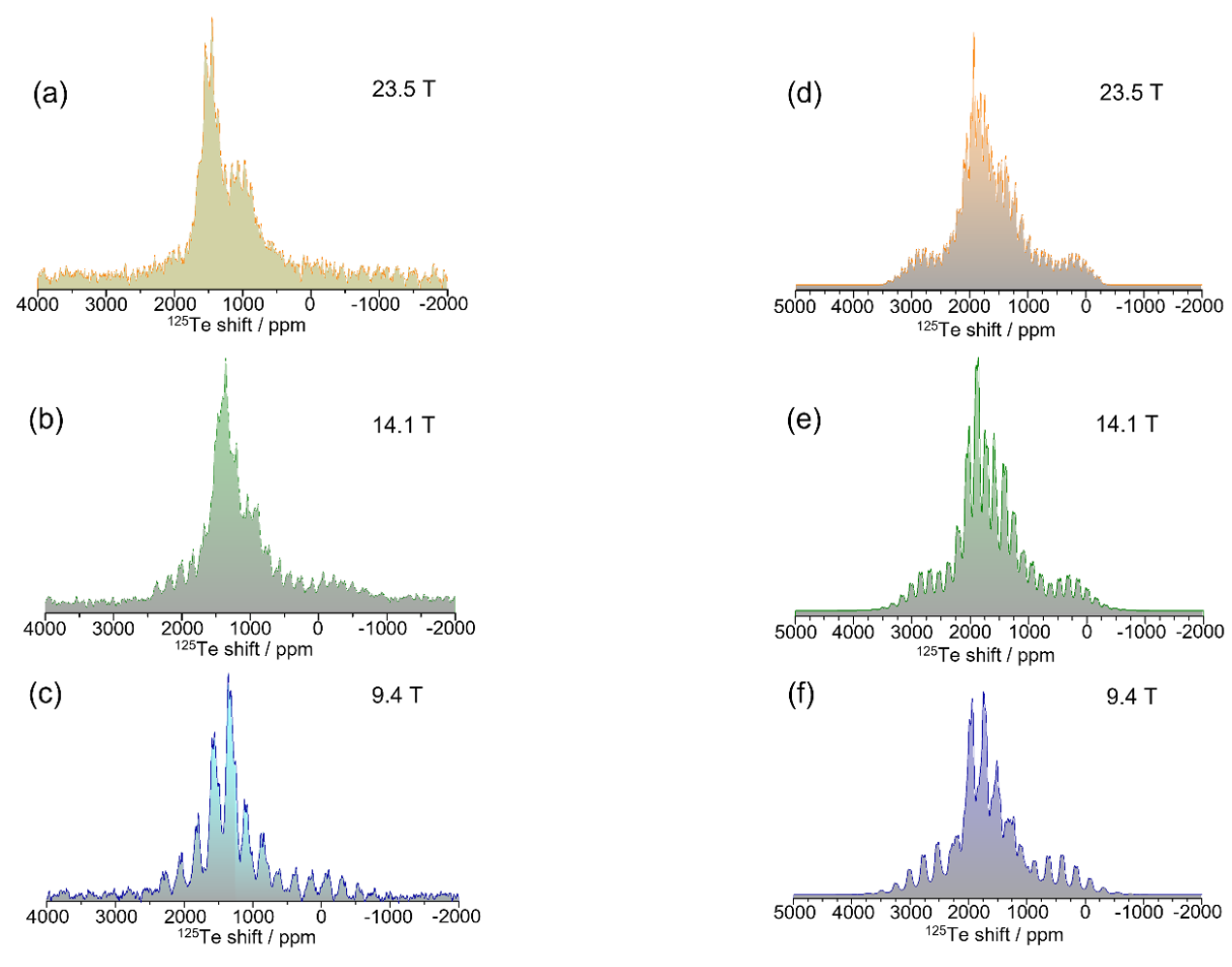}
	\caption{\label{FigS8}Evolution of the $^{125}$Te MAS NMR spectral lineshape for varying magnetic field strength. (a-c) The experimental NMR lineshape for magnetic fields of 23.5 T, 14.1 T, and 9.4 T respectively. (d-f) The calculated NMR lineshape simulated for the same magnetic fields.}
	\end{figure}
No remarkable difference between the experimental and DFT calculated NMR spectra is observed at various magnetic fields, which is indicative of the absence of any major change in the electron energy band structure on increasing the magnetic field, at least in the magnetic field range of 9.4 – 23.5T, and at room temperature. This is further confirmed in the 2-D $^{125}$Te aMAT NMR spectrum at 14.1 T, shown in Supplementary Figure~\ref{FigS9} where no major differences in the evolution of the isotropic shifts are observed at 14.1 T, in comparison to the 9.4 T aMAT spectrum presented in Figure 2a of the main article.  
\begin{figure}[h!]
	\includegraphics[width=0.65\textwidth]{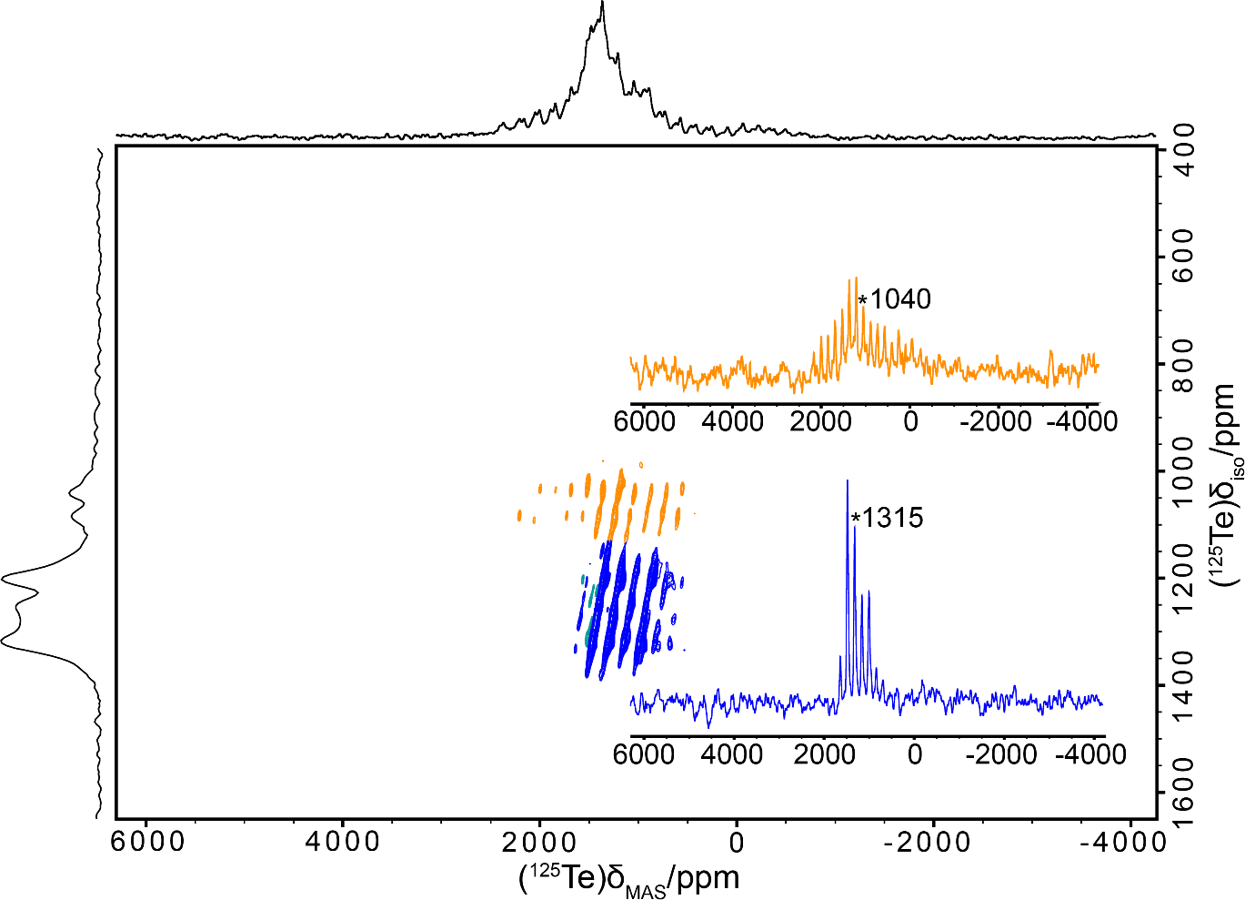}
	\caption{\label{FigS9}The experimental $^{125}$Te MAS NMR aMAT spectrum at 14.1 T and 30 kHz MAS. The isotropic shifts and shift anisotropies of both Te(1),Te(3) and Te(2),Te(4) remain unchanged compared to the aMAT spectrum at 9.4 T.}
	\end{figure}
\FloatBarrier
\section{5.  The evolution of the $\textbf{Te}$ projected Density of States ($\textbf{p}$DOS) close to the Fermi energy.}
\begin{figure}[h!]
	\includegraphics[width=0.9\textwidth]{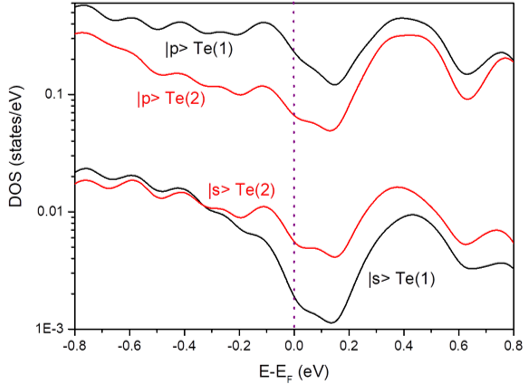}
	\caption{\label{FigS10}Te $\ket{s}$ and $\ket{p}$ electrons pDOS in the vicinity of the Fermi level. The plots show the s-orbital DOS (red) and p-orbital DOS (black) of Te(1) and Te(2) in the presence of SOC. A similar behavior is followed by corresponding pDOS of Te(3) and Te(4).}
	\end{figure}

\begin{figure}[h!]
	\includegraphics[width=0.95\textwidth]{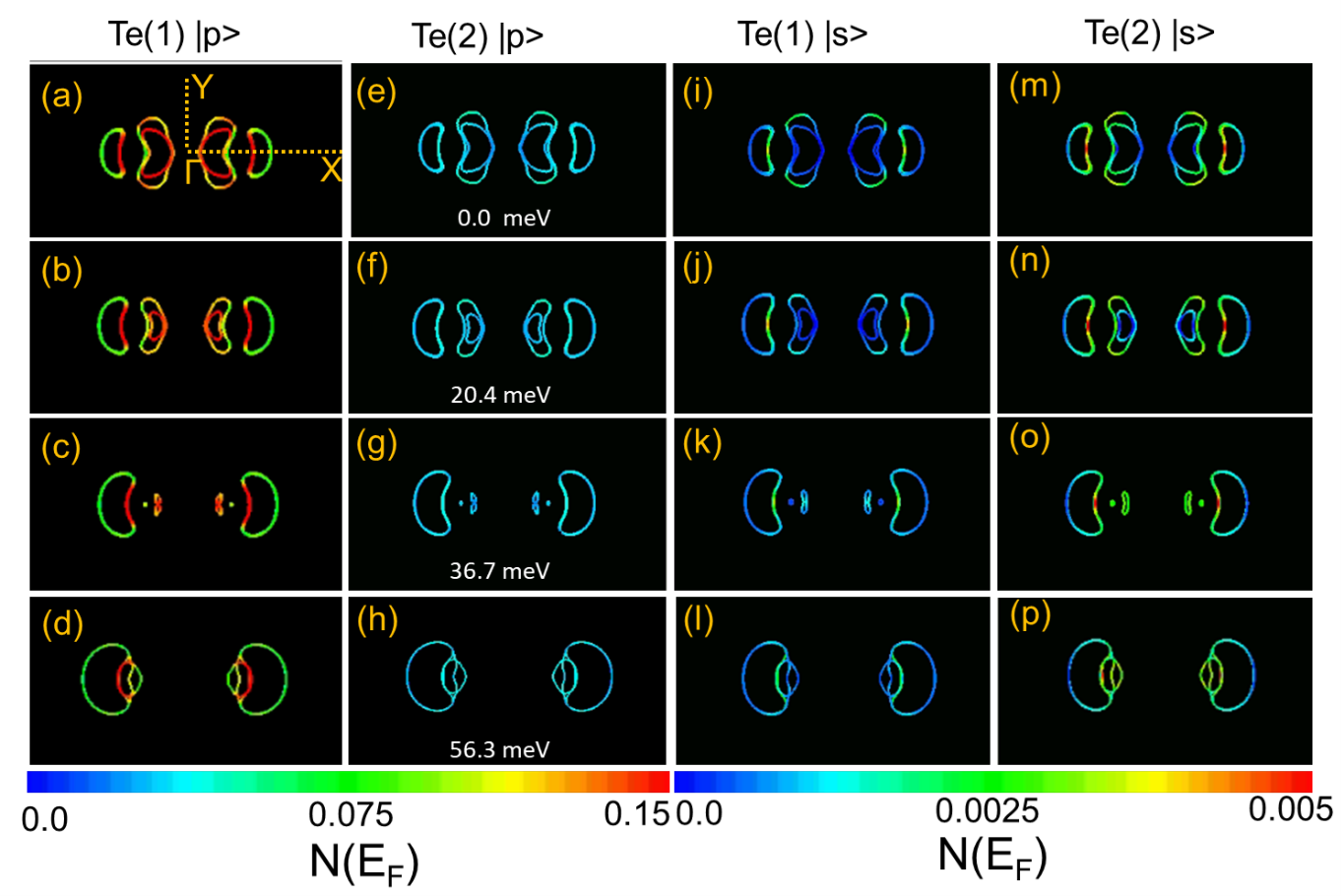}
	\caption{\label{FigS11}k-resolved pDOS of Fermi surface cross sections for s- and p- orbitals at Fermi levels of 0.0, 20.4, 36.7 and 56.3 meV (T$_{d}$ phase of WTe$_{2}$). (a-d) p-orbitals k-pDOS for Te(1) (e-h)  p-orbitals k-pDOS for Te(2). (i-l) s-orbitals k-pDOS for Te(1). (m-p) p-orbitals k-pDOS for Te(2).  The same behavior was observed for Te(3) and Te(4), respectively.}
	\end{figure}

Supplementary Figure~\ref{FigS10} demonstrates the $\ket{s}$ and $\ket{p}$ electrons pDOS for the crystallographically inequivalent Te(1) and Te(2) sites in the vicinity of the Fermi level, calculated withQuantum Espresso. Results corroborate the NMR calculations presented in Supplementary Table 1. The higher $\ket{p}$ electrons pDOS of the Te(1)/Te(3) sites, indicates that p-orbital shielding is the origin of the larger shift anisotropy of the respective NMR signals compared to Te(2)/Te(4), while the higher $\ket{s}$-electron pDOS of Te(2)/Te(4) at the Fermi level explains the larger NMR Fermi contact shielding ($\sigma_{FC}$) they show, compared to the Te(1)/Te(3) sites. 
This is in line with Supplementary Figure~\ref{FigS11}, which shows the FS cross-sections of the s- and p-electron bands at k$_{Z}$=0, and various energy values of the Fermi level, acquired by using the Quantum Espresso and the FermiSurfer software packages \cite{Kawamura2019}. Calculations show that the highest pDOS of the p-orbitals resides in the Te(1) hole pockets (Supplementary Figure~\ref{FigS11}a, whilst the highest s-orbitals pDOS is for the Te(2) sites (Supplementary Figure~\ref{FigS11}m). The results correlate with those in Supplementary Figure~\ref{FigS10} and Supplementary Table 1. Remarkably, upon increasing temperature the Fermi level raises rapidly, and the hole pockets begin to shrink, and finally vanish at E$_{F}$$\sim$40 meV, as also depicted in Figures 3 and 4 of the main article.
The rapid decrease of the FS in WTe$_{2}$ with temperature has been experimentally verified in a number of ARPES and transport properties experiments (refs. 17-20 of the main article). At the same time, a significant increase in the surface of the electron pockets is observed.
Finally, the Fermi velocity distribution at the electron and hole pockets is shown in Supplementary Figure~\ref{FigS12}. Notably, the higher Fermi velocity is observed on the electron pockets; this might explain the steepest slope of the  $\frac{1}{{T{_1}T}}$ vs. $T^{2}$ plot of Te(1) in comparison to Te(2) in Figure 3 of the main article, because  according to ref. \cite{Okvatovity2019}  $\frac{1}{{T{_1}T}}\sim\frac{1}{{\upsilon_F^2}}$.
\begin{figure}[h!]
	\includegraphics[width=1.0\textwidth]{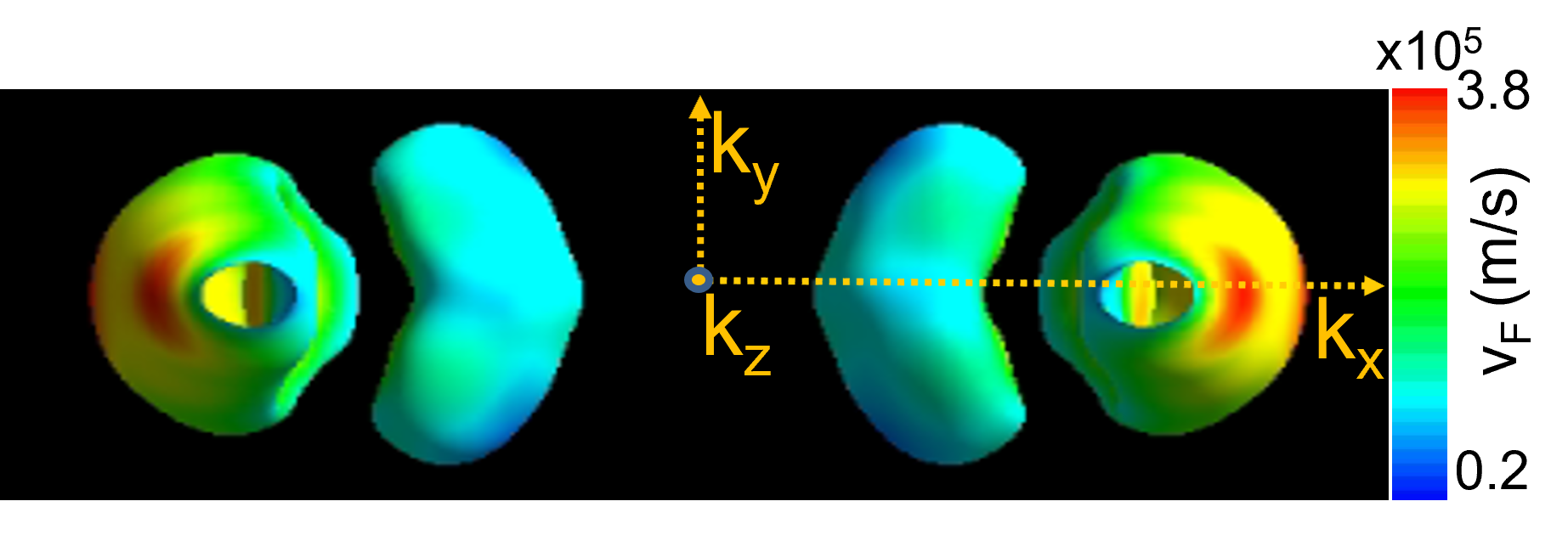}
	\caption{\label{FigS12}THe Fermi velocity map across the hole and the electron pockets in the T$_{d}$ phase of WTe$_{2}$.}
	\end{figure}

\raggedright	
\FloatBarrier
\bibliography{WTe2_supp}